\documentclass[12pt,a4paper]{article}

\usepackage{cite}
\usepackage{graphicx}
\usepackage{epsfig}
\usepackage{amssymb}
\usepackage{amsmath}
\usepackage{latexsym}

\newcommand{\capdef}{}
\newcommand{\mycaption}[2][\capdef]{\renewcommand{\capdef}{#2}%
       \caption[#1]{{\footnotesize #2}}}
\makeatletter
\renewcommand{\fnum@table}{\textbf{\tablename~\thetable}}
\renewcommand{\fnum@figure}{\textbf{\figurename~\thefigure}}
\makeatother

\newcounter{myenumi}

\renewcommand{\themyenumi}{\roman{myenumi}}
{\end{list}}

\newlength{\myem}
\newcounter{mysubequation}[equation]

\makeatletter
\renewcommand{\section}{\@startsection{section}{1}{0em}{-\baselineskip}%
{\baselineskip}{\normalfont\large\bfseries}}
\renewcommand{\subsection}%
{\@startsection{subsection}{2}{0em}{-0.7\baselineskip}%
{0.7\baselineskip}{\normalfont\bfseries}}
\makeatother

\newcommand{\centre}[2]{\multispan{#1}{\hfill #2\hfill}}
\newcommand{\etal}{\textit{et al.}}
\newcommand{\stheta}{\ensuremath{\sin^22\theta_{13}}}
\newcommand{\BB}{$\beta$B}
\newcommand{\sigdm}{\ensuremath{{\rm sign}(\Delta m^2_{31})}}
\newcommand{\delCP}{\ensuremath{\delta_{\rm CP}}}
\newcommand{\thetatt}{\ensuremath{\theta_{23}}}

\def\nue{\ensuremath{\nu_{e}}}
\def\nubare{\ensuremath{\overline{\nu}_{e}}}

\def\numu{\ensuremath{\nu_{\mu}\ }}
\def\nubarmu{\ensuremath{\overline{\nu}_{\mu}}}

\newcommand{\He}{\ensuremath{^6{\mathrm{He}}}}
\newcommand{\Ne}{\ensuremath{^{18}{\mathrm{Ne}}}}

\newcommand{\thetaot}{\ensuremath{\theta_{13}}\,}

\newcommand{\Efin}{E^\text{fin}}
\newcommand{\Emin}{E_\text{min}}

\begin{document}


\renewcommand{\thefootnote}{\alph{footnote}}

\begin{flushright}
LAL-06-35\\
IC/2006/011\\
SISSA 16/2006/EP\\
\end{flushright}

\vspace*{0.5cm}

\renewcommand{\thefootnote}{\fnsymbol{footnote}}
\setcounter{footnote}{-1}

{\begin{center} 
{\Large\textbf{
Physics potential of the CERN--MEMPHYS\\[2mm] 
neutrino oscillation project}
}
\end{center}}

\vspace*{0.5cm}

\begin{center} {\bf
J.-E.\ Campagne$^a$, 
M.\ Maltoni$^b$,
M.\ Mezzetto$^c$, and 
T.\ Schwetz$^d$}
\end{center}

{\it
\begin{center}
  $^a$Laboratoire de l'Acc\'el\'erateur Lin\'eaire,
  Universit\'e Paris-Sud, IN2P3/CNRS\\
  B.P.\ 34, 91898 Orsay Cedex, France\\[2mm]
  $^b$International Centre for Theoretical Physics, 
  Strada Costiera 11, I-31014 Trieste, Italy\\ 
  {\rm and}\\
  Departamento de F\'isica Te\'orica \& Instituto de F\'isica Te\'orica,
  Facultad de Ciencias C-XI,\\ 
  Universidad Aut\'onoma de Madrid, Cantoblanco, E-28049 Madrid, Spain\\[2mm]
  $^c$Istituto Nazionale Fisica Nucleare, Sezione di Padova,
  Via Marzolo 8, 35100 Padova, Italy\\[2mm]
  $^d$Scuola Internazionale Superiore di Studi Avanzati, 
  Via Beirut 2--4, 34014 Trieste, Italy\\ 
  {\rm and}\\
  CERN, Physics Department, Theory Division, CH-1211 Geneva 23, Switzerland
\end{center}}

\vspace*{0.5cm}

\begin{abstract}
We consider the physics potential of CERN based neutrino oscillation
experiments consisting of a Beta Beam (\BB) and a Super Beam (SPL)
sending neutrinos to MEMPHYS, a 440~kt water \v{C}erenkov detector at
Fr\'ejus, at a distance of 130~km from CERN. The $\theta_{13}$
discovery reach and the sensitivity to CP violation are investigated,
including a detailed discussion of parameter degeneracies and
systematical errors. For SPL sensitivities similar to the ones of the
phase~II of the T2K experiment (T2HK) are obtained, whereas the \BB\
may reach significantly better sensitivities, depending on the
achieved number of total ion decays.  The results for the
CERN--MEMPHYS experiments are less affected by systematical
uncertainties than T2HK.
We point out that by a combination of data from \BB\ and SPL a
measurement with antineutrinos is not necessary and hence the same
physics results can be obtained within about half of the measurement time
compared to one single experiment.
Furthermore, it is shown how including data from atmospheric neutrinos in
the MEMPHYS detector allows to resolve parameter degeneracies and, in
particular, provides sensitivity to the neutrino mass hierarchy and
the octant of $\theta_{23}$.
\end{abstract}

\renewcommand{\thefootnote}{\arabic{footnote}}
\setcounter{footnote}{0}

\newpage

\section{Introduction}

In recent years strong evidence for neutrino oscillations has been
obtained in solar~\cite{solar},
atmospheric~\cite{Fukuda:1998mi,Ashie:2005ik},
reactor~\cite{Araki:2004mb}, and accelerator~\cite{Aliu:2004sq}
neutrino experiments. The very near future of long-baseline (LBL)
neutrino experiments is devoted to the study of the oscillation
mechanism in the range of $\Delta m^2_{31} \approx 2.4\times10^{-3} \:
\mathrm{eV}^2$ indicated by atmospheric neutrinos using conventional
$\nu_\mu$ beams.  Similar as in the K2K experiment in
Japan~\cite{Aliu:2004sq}, the presently running MINOS experiment in
the USA~\cite{MINOS} uses a low energy beam to measure $\Delta
m^2_{31}$ by observing the $\nu_\mu\rightarrow\nu_\mu$ disappearance
probability, while the OPERA~\cite{OPERA} experiment will be able to
detect $\nu_\tau$ appearance within the high energy CERN--Gran Sasso
beam~\cite{CNGS}.
If we do not consider the LSND anomaly~\cite{LSND} that will be
further studied soon by the MiniBooNE experiment~\cite{MINIBOONE}, all
data can be accommodated within the three flavor scenario (see
Refs.~\cite{FOGLILISI05,Maltoni:2004ei} for recent global analyses),
and neutrino oscillations are described by two neutrino mass-squared
differences ($\Delta m^2_{21}$ and $\Delta m^2_{31}$) and the $3\times
3$ unitary Pontecorvo-Maki-Nakagawa-Sakata (PMNS) lepton mixing
matrix~\cite{PMNS} with three angles
($\theta_{12}$,$\theta_{13}$,$\theta_{23}$) and one Dirac CP phase
$\delCP$.

Future tasks of neutrino physics are an improved sensitivity to the
last unknown mixing angle, $\theta_{13}$, to explore the CP violation
mechanism in the leptonic sector, and to determine the sign of $\Delta
m^2_{31}$ which describes the type of the neutrino mass hierarchy
(normal, $\Delta m^2_{31} > 0$ or inverted, $\Delta m^2_{31} < 0$).
The present upper bound on $\theta_{13}$ is dominated by the
constraint from the Chooz reactor experiment~\cite{CHOOZ}. A global
analysis of all data yields $\sin^22\theta_{13}<0.082$ at
90\%~CL~\cite{Maltoni:2004ei}. A main purpose of upcoming reactor and
accelerator experiments is to improve this bound or to reveal a finite
value of $\theta_{13}$. In reactor experiments, one uses $\bar{\nu}_e$
in disappearance mode and the sensitivity is increased with respect to
present experiments by the use of a near detector close to the
reactor~\cite{Wpaper}. In accelerator experiments, the first
generation of so-called Super Beams with sub-mega watt proton drivers
such as T2K (phase-I)~\cite{T2K} and NO$\nu$A~\cite{Ayres:2004js}, the
appearance channel $\nu_\mu\to\nu_e$ is explored. This next generation
of reactor and Super Beam experiments will reach sensitivities of the
order of $\sin^22\theta_{13} \lesssim 0.01$ ($90\%$~CL) within a time
scale of several years~\cite{Huber:2003pm}.
Beyond this medium term program, there are several projects on how to
enter the high precision age in neutrino oscillations and to attack
the ultimate goals like the discovery of leptonic CP violation or the
determination of the neutrino mass hierarchy. In accelerator
experiments, one can extend the Super Beam concept by moving to
multi-mega watt proton drivers~\cite{T2K,Albrow:2005kw,SPL,BNLHS} or
apply novel technologies, such as neutrino beams from decaying ions
(so-called Beta Beams)~\cite{zucchelli,Albright:2004iw} or from
decaying muons (so-called Neutrino
Factories)~\cite{Albright:2004iw,Blondel:2004ae}.

In this work we focus on possible future neutrino oscillation
facilities hosted at CERN, namely a multi-mega watt Super Beam
experiment based on a Super Proton Linac (SPL)~\cite{Campagne:2004wt}
and a $\gamma = 100$ Beta Beam
(\BB)~\cite{Mezzetto:2003ub}. These experiments will search for
$\stackrel{\scriptscriptstyle (-)}{\nu}_\mu \to
\stackrel{\scriptscriptstyle(-)}{\nu}_e$ and
$\stackrel{\scriptscriptstyle (-)}{\nu}_e \to
\stackrel{\scriptscriptstyle(-)}{\nu}_\mu$ appearance, respectively,
by sending the neutrinos to a mega ton scale water \v{C}erenkov
detector (MEMPHYS)~\cite{memphys}, located at a distance of 130~km from
CERN under the Fr\'ejus mountain. Similar detectors are under
consideration also in the US (UNO~\cite{UNO}) and in Japan
(Hyper-K~\cite{T2K,Nakamura:2003hk}).
We perform a detailed analysis of the SPL and \BB\ physics
potential, discussing the discovery reach for $\theta_{13}$ and
leptonic CP violation. In addition we consider the possibility to
resolve parameter degeneracies in the LBL data by using the
atmospheric neutrinos available in the mega ton
detector~\cite{Huber:2005ep}. This leads to a sensitivity to the
neutrino mass hierarchy of the CERN--MEMPHYS experiments, despite the
rather short baseline.
The physics performances of \BB\ and SPL are compared to the ones
obtainable at the second phase of the T2K experiment in Japan, which
is based on an upgraded version of the original T2K beam and the
Hyper-K detector (T2HK)~\cite{T2K}.

The outline of the paper is as follows.  In Sec.~\ref{sec:analysis} we
summarize the main characteristics of the \BB, SPL, and T2HK
experiments and give general details of the physics analysis methods,
whereas in Sec.~\ref{sec:experiments} we describe in some detail the
MEMPHYS detector, the \BB, the SPL Super Beam, and our atmospheric
neutrino analysis. In Sec.~\ref{sec:degeneracies} we review the
problem of parameter degeneracies and discuss its implications for the
experiments under consideration. In Sec.~\ref{sec:sensitivities} we
present the sensitivities to the ``atmospheric parameters''
$\theta_{23}$ and $\Delta m^2_{31}$, the $\theta_{13}$ discovery
potential, and the sensitivity to CP violation. We also investigate in
some detail the impact of systematical errors. In
Sec.~\ref{sec:synergies} we discuss synergies which are offered by the
CERN--MEMPHYS facilities.  We point out advantages of the case when
\BB\ and SPL are available simultaneously, and we consider the use of
atmospheric neutrino data in MEMPHYS in combination with the LBL
experiments. Our results are summarized in Sec.~\ref{sec:conclusions}.

\section{Experiments overview and analysis methods}
\label{sec:analysis}

In this section we give the most important experimental parameters
which we adopt for the simulation of the CERN--MEMPHYS experiments
\BB\ and SPL, as well as for the T2HK experiment in Japan. These
parameters are summarized in Tab.~\ref{tab:setups}. For all
experiments the detector mass is 440~kt, and the running time is 10
years, with a division in neutrino and antineutrino running time in
such a way that roughly an equal number of events is obtained. We
always use the total available information from appearance as well as
disappearance channels including the energy spectrum. For all three
experiments we adopt rather optimistic values for the systematical
uncertainties of 2\% as default values, but we also consider the case
when systematics are increased to 5\%. These errors are uncorrelated
between the various signal channels (neutrinos and antineutrinos), and
between signals and backgrounds.

\begin{table}
  \centering
  \begin{tabular}{lcc@{\qquad\qquad}c}
  \hline\noalign{\smallskip}
       & \BB & SPL & T2HK \\
  \noalign{\smallskip}\hline\noalign{\smallskip}
  Detector mass & 440~kt & 440~kt & 440~kt\\
  Baseline      & 130 km & 130 km & 295 km \\
  Running time ($\nu + \bar\nu$) 
                & 5 + 5 yr & 2 + 8 yr & 2 + 8 yr \\
  Beam intensity  & $5.8\,(2.2) \cdot 10^{18}$ He (Ne) dcys/yr & 4 MW & 4 MW\\
  Systematics on signal  & 2\% & 2\% & 2\%\\
  Systematics on backgr. & 2\% & 2\% & 2\%\\
  \noalign{\smallskip}\hline
  \end{tabular}
  \mycaption{Summary of default parameters used for the simulation of the
  \BB, SPL, and T2HK experiments.\label{tab:setups}}
\end{table}

A more detailed description of the CERN--MEMPHYS experiments is given
in Sec.~\ref{sec:experiments}. For the T2HK simulation we use the
setup provided by GLoBES~\cite{Globes} based on
Ref.~\cite{Huber:2002mx}, which follows closely the LOI~\cite{T2K}. In
order to allow a fair comparison we introduce the following changes
with respect to the configuration used in Ref.~\cite{Huber:2002mx}:
The fiducial mass is set to 440~kt, the systematical errors on the
background and on the $\nu_e$ and $\bar\nu_e$ appearance signals is
set to 2\%, and we use a total running time of 10 years, divided into
2 years of data taking with neutrinos and 8 years with
antineutrinos. We include an additional background from the
$\bar\nu_\mu \to \bar\nu_e$ ($\nu_\mu \to \nu_e$) channel in the
neutrino (antineutrino) mode. Furthermore, we use
the same CC detection cross section as for the \BB/SPL
analysis~\cite{Nuance}. For more details see
Refs.~\cite{T2K,Huber:2002mx}.

Recently the idea was put forward to observe the T2K beam with a
second detector placed in
Korea~\cite{Ishitsuka:2005qi,Kajita:2006bt,Hagiwara:2005pe}.  In the
so-called T2KK setup the Hyper-Kamiokande detector is split into two
detectors of 270~kt each, one of them is located in Korea at a
distance of about 1050~km from the source, and the other is placed at
Kamioka at a distance of 295~km. Here we will confine ourselves to the
standard T2HK setup, since the purpose of our work is not a T2HK
optimization study investigating various configurations for that
experiment. In contrast, here T2HK mainly serves as a point of
reference to which we compare the CERN--MEMPHYS experiments. For this
aim we prefer to stick to the ``minimal'' one-detector configuration
at a relatively short baseline, since two-detector setups with very
long baselines clearly represent a different class of experiments
whose consideration goes beyond the scope of the present
work. Nevertheless we will comment briefly also on T2KK performances
obtained in Refs.~\cite{Ishitsuka:2005qi,Kajita:2006bt}.

\begin{table}
  \centering
  \begin{tabular}{lcccccc}
  \hline\noalign{\smallskip}
       & \centre{2}{\BB} & \centre{2}{SPL} & \centre{2}{T2HK} \\
  \noalign{\smallskip}\hline\noalign{\smallskip}
  & $\delCP=0$ & $\delCP=\pi/2$ & $\delCP=0$ & $\delCP=\pi/2$ & $\delCP=0$ & $\delCP=\pi/2$\\
  \noalign{\smallskip}\hline\noalign{\smallskip}
  appearance $\nu$ & & & & & & \\
  background       & \centre{2}{143} &\centre{2}{622} &\centre{2}{898}\\
  $\stheta=0$      & \centre{2}{28}  &\centre{2}{51}  &\centre{2}{83}  \\
  $\stheta=10^{-3}$&    76  &   88   &   105  &   14  &   178 &    17  \\ 
  $\stheta=10^{-2}$&   326  &  365   &   423  &  137  &   746 &   238  \\
  \noalign{\smallskip}\hline\noalign{\smallskip}
  appearance $\bar\nu$ & & & & & & \\
  background       & \centre{2}{157} &\centre{2}{640} &\centre{2}{1510}\\
  $\stheta=0$      & \centre{2}{31}  &\centre{2}{57}  &\centre{2}{93}  \\
  $\stheta=10^{-3}$&    83  &   12   &   102  &  146  &   192 &   269  \\ 
  $\stheta=10^{-2}$&   351  &  126   &   376  &  516  &   762 &  1007  \\
  \noalign{\smallskip}\hline\noalign{\smallskip}
  disapp. $\nu$ &\centre{2}{100315}&\centre{2}{21653}&\centre{2}{24949}\\
  background    & \centre{2}{6}   &\centre{2}{1}    &\centre{2}{444}\\
  disapp. $\bar\nu$&\centre{2}{84125}&\centre{2}{18321}&\centre{2}{34650}\\
  background       &\centre{2}{5}    &\centre{2}{1}    &\centre{2}{725}\\
  \noalign{\smallskip}\hline
  \end{tabular}
  \mycaption{Number of events for appearance and disappearance signals
  and backgrounds for the \BB, SPL, and T2HK experiments as
  defined in Tab.~\ref{tab:setups}. For the appearance signals the
  event numbers are given for several values of $\stheta$ and $\delCP
  = 0$ and $\pi/2$. The background as well as the disappearance event
  numbers correspond to $\theta_{13}=0$. For the other oscillation
  parameters the values of Eq.~(\ref{eq:default-params}) are
  used.\label{tab:events}}
\end{table}

In Tab.~\ref{tab:events} we give the number of signal and background
events for the experiment setups as defined in Tab.~\ref{tab:setups}.
For the appearance channels ($\stackrel{\scriptscriptstyle (-)}{\nu}_e
\to \stackrel{\scriptscriptstyle(-)}{\nu}_\mu$ for the \BB\ and
$\stackrel{\scriptscriptstyle (-)}{\nu}_\mu \to
\stackrel{\scriptscriptstyle(-)}{\nu}_e$ for SPL and T2HK) we give the
signal events for various values of $\theta_{13}$ and $\delCP$. The
``signal'' events for $\theta_{13} = 0$ are appearance events induced by
the oscillations with $\Delta m^2_{21}$. The value $\stheta = 10^{-3}$
corresponds roughly to the sensitivity limit for the considered
experiments, whereas $\stheta = 10^{-2}$ gives a good sensitivity
to CP violation. This can be appreciated by comparing the values of
$\nu$ and $\bar\nu$ appearance events for $\delCP = 0$ and $\pi/2$. In
the table the background to the appearance signal is given for
$\theta_{13} = 0$. Note that in general the number of background
events depends also on the oscillation parameters, since also the
background neutrinos in the beam oscillate. This effect is
consistently taken into account in the analysis, however, for the
parameter values in the table the change in the background events due
to oscillations is only of the order of a few events.

The physics analysis is performed with the GLoBES open source
software~\cite{Globes}, which provides a convenient tool to simulate
long-baseline experiments and compare different facilities in a
unified framework. The experiment definition (AEDL) files for the \BB\
and SPL simulation with GLoBES are available at Ref.~\cite{Globes}.
In the analysis parameter degeneracies and correlations are fully
taken into account and in general all oscillation parameters are
varied in the fit.
To simulate the ``data'' we adopt the following
set of ``true values'' for the oscillation parameters:
\begin{equation}\label{eq:default-params}
\begin{array}{l@{\qquad}l}
  \Delta m^2_{31} = +2.4 \times 10^{-3}~\mathrm{eV}^2\,, & 
  \sin^2\theta_{23} = 0.5\,,\\ 
  \Delta m^2_{21} = 7.9 \times 10^{-5}~\mathrm{eV}^2 \,,&
  \sin^2\theta_{12} = 0.3 \,,
\end{array}
\end{equation} 
and we include a prior knowledge of these values with an accuracy of
10\% for $\theta_{12}$, $\theta_{23}$, $\Delta m^2_{31}$, and 4\% for
$\Delta m^2_{21}$ at 1$\sigma$. These values and accuracies are
motivated by recent global fits to neutrino oscillation
data~\cite{FOGLILISI05,Maltoni:2004ei}, and they are always used
except where explicitly stated otherwise.

\section{The CERN--MEMPHYS experiments}
\label{sec:experiments}

\subsection{The MEMPHYS detector}

\begin{figure}
\centering
\includegraphics[width=0.65\textwidth]{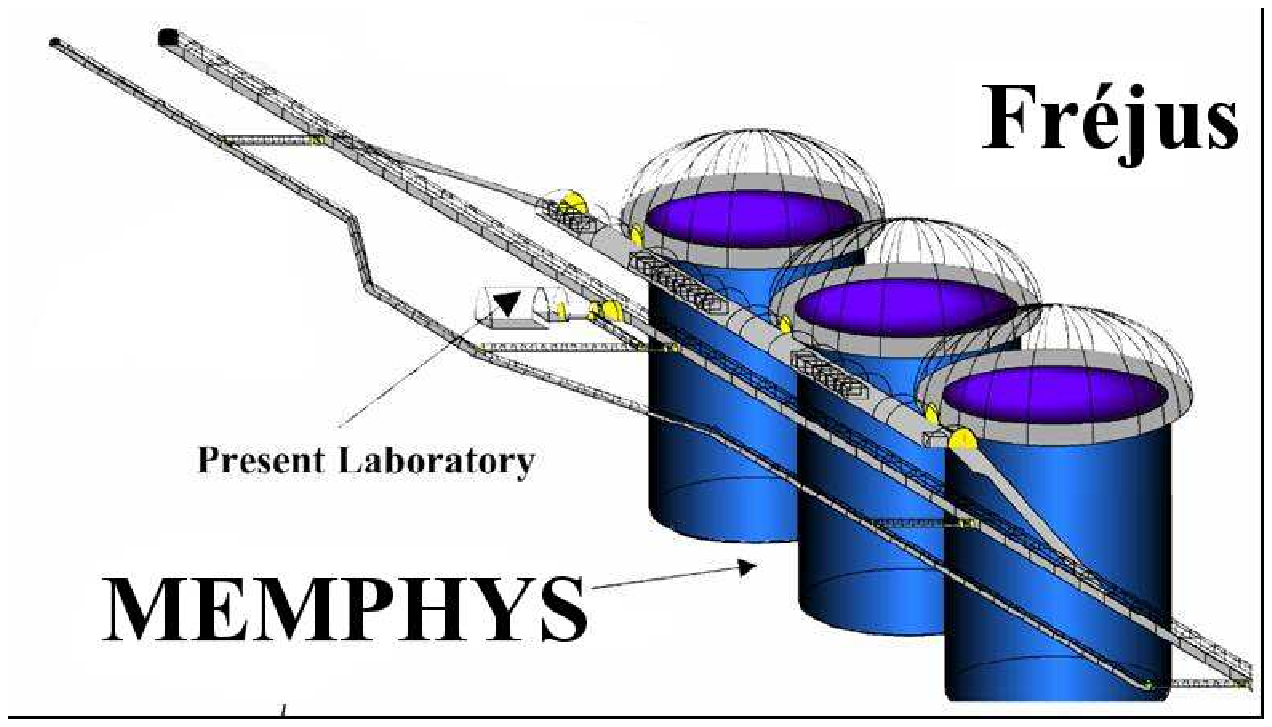}
\mycaption{\label{fig:MEMPHYS}Sketch of the MEMPHYS detector under the
Fr\'ejus mountain.}	
\end{figure}

MEMPHYS (MEgaton Mass PHYSics)~\cite{memphys} is a mega ton class
water \v{C}erenkov detector in the straight extrapolation of
Super-Kamiokande, located at Fr\'ejus, at a distance of 130~km from
CERN. It is an alternative design of the UNO~\cite{UNO} and
Hyper-Kamiokande~\cite{Nakamura:2003hk} detectors and shares the same
physics case, both from the non-accelerator domain (nucleon decay,
super nova neutrino detection, solar neutrinos, atmospheric neutrinos)
and from the accelerator domain which is the subject of this paper. A
recent civil engineering pre-study to envisage the possibly of large
cavity excavation located under the Fr\'ejus mountain (4800~m.e.w.)
near the present Modane underground laboratory has been undertaken.
The main result of this pre-study is that MEMPHYS may be built with
present techniques as a modular detector consisting of several shafts,
each with 65~m in diameter, 65~m in height for the total water
containment. A schematic view of the layout is shown in
Fig.~\ref{fig:MEMPHYS}. For the present study we have chosen a
fiducial mass of 440~kt which means 3 shafts and an inner detector of
57~m in diameter and 57~m in height.  Each inner detector may be
equipped with photo detectors (81000 per shaft) with a 30\%
geometrical coverage and the same photo-statistics as Super-Kamiokande
(with a 40\% coverage). In principle up to 5 shafts are possible,
corresponding to a fiducial mass of 730~kt.
The Fr\'ejus site offers a natural protection against cosmic rays by a
factor $10^6$. If not mentioned otherwise, the event selection and
particle identification are the Super-Kamiokande algorithms results.

\subsection{The $\gamma = 100\times100$ baseline Beta Beam}

The concept of a Beta Beam (\BB) has been introduced by P.~Zucchelli
in Ref.~\cite{zucchelli}. Neutrinos are produced by the decay of
radioactive isotopes which are stored in a decay ring. An important
parameter is the relativistic gamma factor of the ions, which
determines the energy of the emitted neutrinos. \BB\ performances have
been computed previously for $\gamma(\He)= 66$~\cite{Mezzetto:2003ub},
100~\cite{MyNufact04,Donini:2004hu,JJHigh2}, 150~\cite{JJHigh2},
200~\cite{LindnerBB}, 350~\cite{JJHigh2},
500~\cite{JJHigh1,LindnerBB}, 1000~\cite{LindnerBB},
2000~\cite{JJHigh1}, 2488~\cite{Terranova}. Reviews can be found in
Ref.~\cite{BB-Reviews}, the physics potential of a very low gamma \BB\
has been studied in Ref.~\cite{Volpe}. Performances of a \BB\ with
$\gamma > 150$ are extremely promising, however, they are neither
based on an existing accelerator complex nor on detailed calculations
of the ion decay rates. For a CERN based \BB, fluxes have been
estimated in Ref.~\cite{Lindroos} and a design study is in progress
for the facility \cite{Eurisol}. In this work we assume an integrated
flux of neutrinos in 10 years corresponding to $2.9\cdot 10^{19}$
useful \He\ decays and $1.1 \cdot 10^{19}$ useful \Ne\ decays. These
fluxes have been assumed in all the physics papers quoted above, and
they are two times higher than the baseline fluxes computed in
Ref.~\cite{Lindroos}. These latter fluxes suffer for the known
limitations of the PS and SPS synchrotrons at CERN, ways to improve
them have been delineated in Ref.~\cite{Lindroos-Optimization}.

The infrastructure available at CERN as well as the MEMPHYS
location at a distance of 130~km suggest a $\gamma$-factor of about
$100$. Such a value implies a mean neutrino energy of 400~MeV, which
leads to the oscillation maximum at about 200~km for $\Delta m^2_{31}
= 2.4\times 10^{-3}$~eV$^2$.  We have checked that the performance at
the somewhat shorter baseline of 130~km is rather similar to the one
at the oscillation maximum. Moreover, the purpose of this paper is to
estimate the physics potential for a realistic set-up and not to study
the optimization of the \BB\ regardless of any logistic consideration
(see, e.g., Refs.~\cite{LindnerBB,JJHigh2} for such optimization
studies).

\begin{figure}[!t]
  \centering
  \includegraphics[width=0.65\textwidth]{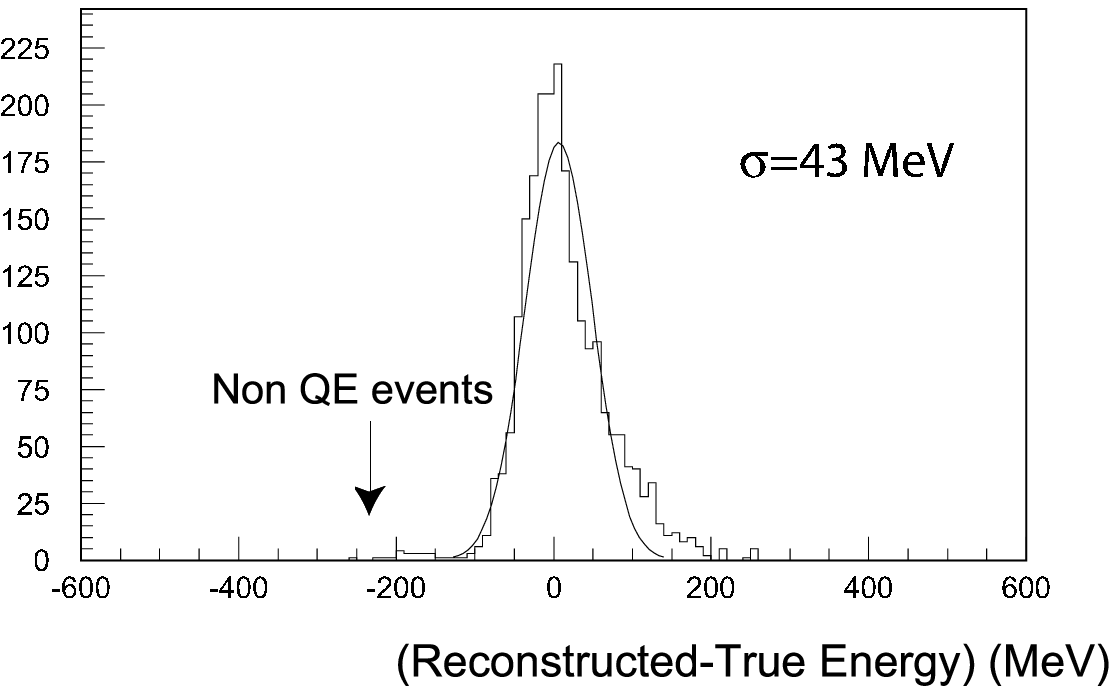}
  \mycaption{\label{fig:QE-Energy} Energy resolution for \nue\
  interactions in the 200--300~MeV energy range. The quantity
  displayed is the difference between the reconstructed and the true
  neutrino energy.}
\end{figure}

The signal events from the $\nu_e \to \nu_\mu$ neutrino and
antineutrino appearance channels in the \BB\ are \numu charged current
(CC) events. The Nuance v3r503 Monte Carlo code~\cite{Nuance} is used
to generate signal events. The selection for these events is based on
standard Super-Kamiokande particle identification algorithms.  The
muon identification is reinforced by asking for the detection of the
Michel decay electron. 
The neutrino energy is reconstructed by smearing momentum and
direction of the charged lepton with the Super-Kamiokande resolution
functions, and applying quasi-elastic (QE) kinematics assuming the
known incoming neutrino direction. Energy reconstruction in the \BB\
energy range is remarkably powerful, and the contamination of non-QE
events very small, as shown in Fig.~\ref{fig:QE-Energy}.
As pointed out in Ref.~\cite{JJHigh2}, it is necessary to use a
migration matrix for the neutrino energy reconstruction to properly
handle Fermi motion smearing and the non-QE event contamination.  We
use 100~MeV bins for the reconstructed energy and 40~MeV bins for the
true neutrino energy.  Four migration matrices (for
$\nu_e,\bar\nu_e,\nu_\mu,\bar\nu_\mu$) are applied to signal events as
well as backgrounds. As suggested from Fig.~\ref{fig:QE-Energy} the 
results using migration matrices are very similar to a Gaussian
energy resolution.

Backgrounds from charged pions and atmospheric neutrinos are computed
with the identical analysis chain as signal events.
Charged pions generated in NC events (or in NC-like events where the
leading electron goes undetected) are the main source of background for
the experiment. To compute this background inclusive NC and CC events
have been generated with the \BB\ spectrum. Events have been selected
where the only visible track is a charged pion above the \v{C}erenkov
threshold. Particle identification efficiencies have been applied to
those particles. The probability for a pion to survive in water until
its decay has been computed with Geant~3.21 and cross-checked with a
Fluka~2003 simulation. This probability is different for positive and
negative pions, the latter having a higher probability to be absorbed
before decaying. The surviving events are background, and the
reconstructed neutrino energy is computed misidentifying these pions
as muons. Event rates are reported in Tab.~\ref{tab:sigbck}. From
these numbers it becomes evident that requiring the detection of the
Michel electron provides an efficient cut to eliminate the pion
background.
These background rates are significantly smaller than quoted in
Ref.~\cite{MyNufact04}, where pion decays were computed with the
same probabilities as for muons and they are slightly different
from those quoted in Ref.~\cite{ MezzettoNuFact05}, where an
older version of Nuance had been used.
The numbers of Tab.~\ref{tab:sigbck} have been cross-checked by
comparing the Nuance and Neugen~\cite{Neugen} event
generators, finding a fair agreement in background rates and energy shape.

\begin{table}[t]
     \centering
     \begin{tabular}{l@{\qquad}rrr@{\qquad}rrr}
     \hline\noalign{\smallskip}
       & \multicolumn{3}{ c }{\Ne} & \multicolumn{3}{c}{\He} \\
     \hline\noalign{\smallskip}
       & \numu CC & $\pi^+$ & $\pi^-$ & \nubarmu CC & $\pi^+$ & $\pi^-$ \\
     \hline\noalign{\smallskip}
      Generated ev.\ & 115367   &  557   &  341 & 101899 &  674   &  400 \\
      Particle ID    &  95717   &  204   &  100 & 85285  &  240   &  118 \\
      Decay          &  61347   &  107   &    8 & 69242  &  120   &    8 \\
\hline\noalign{\smallskip}
    \end{tabular}
    \mycaption{\label{tab:sigbck} Events for the \BB\ in a 4400~kt~yr
    exposure.  \numu(\nubarmu) CC events are computed assuming full
    oscillations ($P_{\nu_e\to\nu_\mu} = 1$), and pion backgrounds are
    computed from \nue(\nubare) CC+NC events. In the rows we give the
    number events generated within the fiducial volume (``Generated
    ev.''), after muon particle identification (``Particle ID''), and
    after applying a further identification requiring the detection of
    the Michel electron (``Decay''). }
\end{table}

Also atmospheric neutrinos can constitute an important source of
background~\cite{zucchelli,JJHigh2,JJHigh1,MezzettoNuFact05}. This
background can be suppressed only by keeping a very short duty cycle
($2.2 \cdot 10^{-3}$ is the target value for the \BB\ design study),
and this in turn is one of the most challenging bounds on the design
of the Beta Beam complex. Following Ref.~\cite{MezzettoNuFact05} we
include the atmospheric neutrino background based on a Monte Carlo
simulation using Nuance. Events are reconstructed as if they were
signal neutrino events. We estimate that 5 events/year would survive
the analysis chain in a full solar year (the \BB\ should run for about
1/3 of this period) and include these events as backgrounds in the
analysis. Under these circumstances, the present value of the \BB\
duty cycle seems to be an overkill, it could be reduced by a factor 5
at least, see also Ref.~\cite{MezzettoNuFact05} for a discussion of
the effect of a higher duty cycle.

\subsection{The $3.5$-GeV SPL Super Beam}

In the recent Conceptual Design Report 2 (CDR2) the foreseen Super
Proton Linac (SPL)~\cite{SPL} will provide the protons for the muon
production in the context of a Neutrino Factory, and at a first stage
will feed protons to a fixed target experiment to produce an intense
conventional neutrino beam (``Super Beam''). The parameters of the
beam line take into account the optimization~\cite{Campagne:2004wt} of
the beam energy as well as the secondary particle focusing and decay
to search for $\nu_\mu \rightarrow \nu_e$ and $\bar{\nu}_\mu
\rightarrow \bar{\nu}_e$ appearance as well as $\nu_\mu$,
$\bar\nu_\mu$ disappearance in a mega ton scale water \v{C}erenkov
detector. In particular, a full simulation of the beam line from the
proton on target interaction up to the secondary particle decay tunnel
has been performed. The proton on a liquid mercury target (30~cm long,
$7.5$~mm radius, 13.546 density) has been simulated with
FLUKA~2002.4~\cite{FLUKA} while the horn focusing system and the decay
tunnel simulation has been preformed with
GEANT~3.21~\cite{GEANT}.\footnote{Although there are differences
between the predicted pion and kaon productions as a function of
proton kinetic energy with FLUKA~2002.4 and 2005.6, the results are
consistent for the relevant energy of 3.5~GeV. We emphasize that the
pion and the kaon production cross-sections are waiting for
experimental confirmation~\cite{HARP-MINERVA} and a new optimization
would be required if their is a disagreement with the present
knowledge.}

\begin{figure}[!t]
  \centering 
  \includegraphics[width=0.65\textwidth]{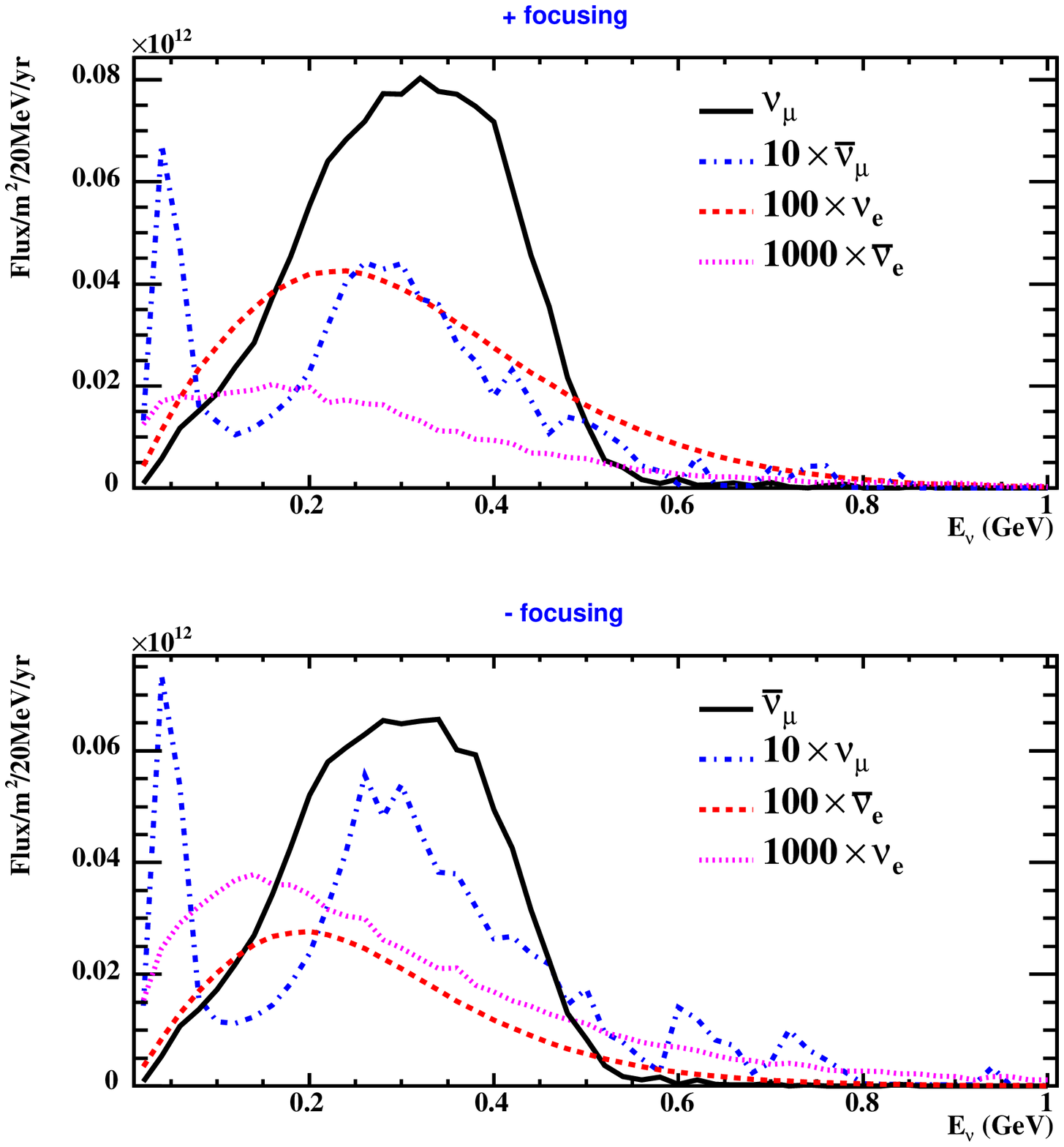}
  \mycaption{\label{fig:fluxSPLContrib} Neutrino fluxes, at $130$~km
  from the target with the horns focusing the positive particles
  (top panel) or the negative particles (bottom panel). The fluxes are
  computed for a SPL proton beam of $3.5$~GeV (4~MW), a decay tunnel
  with a length of $40$~m and a radius of $2$~m.}
\end{figure}

Since the optimization requirements for a Neutrino Factory are rather
different than for a Super Beam the new SPL configuration has a
significant impact on the physics performance (see
Ref.~\cite{Campagne:2004wt} for a detailed discussion).  The SPL
fluxes of the four neutrino species ($\nu_\mu$, $\nu_e$,
$\bar{\nu}_\mu$, $\bar{\nu}_e$) for the positive ($\nu_\mu$ beam) and
the negative focusing ($\bar{\nu}_\mu$ beam) are show in
Fig.~\ref{fig:fluxSPLContrib}.  The total number of $\nu_\mu$
($\bar{\nu}_\mu$) in positive (negative) focusing is about
$1.18\,(0.97) \times 10^{12}\:\mathrm{m}^{-2}\mathrm{y}^{-1}$ with an
average energy of $300$~MeV. The $\nu_e$ ($\bar{\nu}_e$) contamination
in the $\nu_\mu$ ($\bar\nu_\mu$) beam is around $0.7\%$
($6.0\%$). 
Following Ref.~\cite{Mezzetto:2003mm}, the $\pi^o$ background is
reduced using a tighter PID cut compared to the standard
Super-Kamiokande analysis used in K2K, but the cuts are looser than
for T2K. Indeed, at SPL energies the $\pi^o$ background is less severe
than for T2HK. This is because the resonant cross section is
suppressed, and the produced pions have an energy where the angle
between the two gammas is very wide, leading to a small probability
that the two gamma rings overlap. This results in a higher signal
efficiency of SPL compared to T2HK (70\% against 40\%) and a smaller
rate of $\pi^o$ background.
The Michel electron is required for the $\mu$ identification.
For the $\nu_\mu \rightarrow \nu_e$ channel the background consists
roughly of 90\% $\nu_e \rightarrow \nu_e$ CC interactions, 6\% $\pi^o$
from NC interactions, 3\% miss identified muons from $\nu_\mu
\rightarrow \nu_\mu$ CC, and 1\% $\bar{\nu}_e \rightarrow \bar{\nu}_e$
CC interactions. For the $\bar{\nu}_\mu \rightarrow \bar{\nu}_e$
channel the contributions to the background are 45\% $\bar{\nu}_e
\rightarrow \bar{\nu}_e$ CC interactions, 35\% $\nu_e \rightarrow
\nu_e$ CC interactions, 18\% $\pi^o$ from NC interactions and 2\% miss
identified muons from $\bar{\nu}_\mu \rightarrow \bar{\nu}_\mu$ CC.
In addition we include the events from the contamination of 
``wrong sign'' muon-neutrinos due to $\bar\nu_\mu \to \bar\nu_e$
($\nu_\mu \to \nu_e$) oscillations in the neutrino (antineutrino)
mode. 
We have checked that with the envisaged duty cycle of $2.4\times
10^{-4}$ the background from atmospheric neutrinos is negligible for
the SPL.

\begin{figure}[!t]
  \centering
  \includegraphics[width=0.5\textwidth]{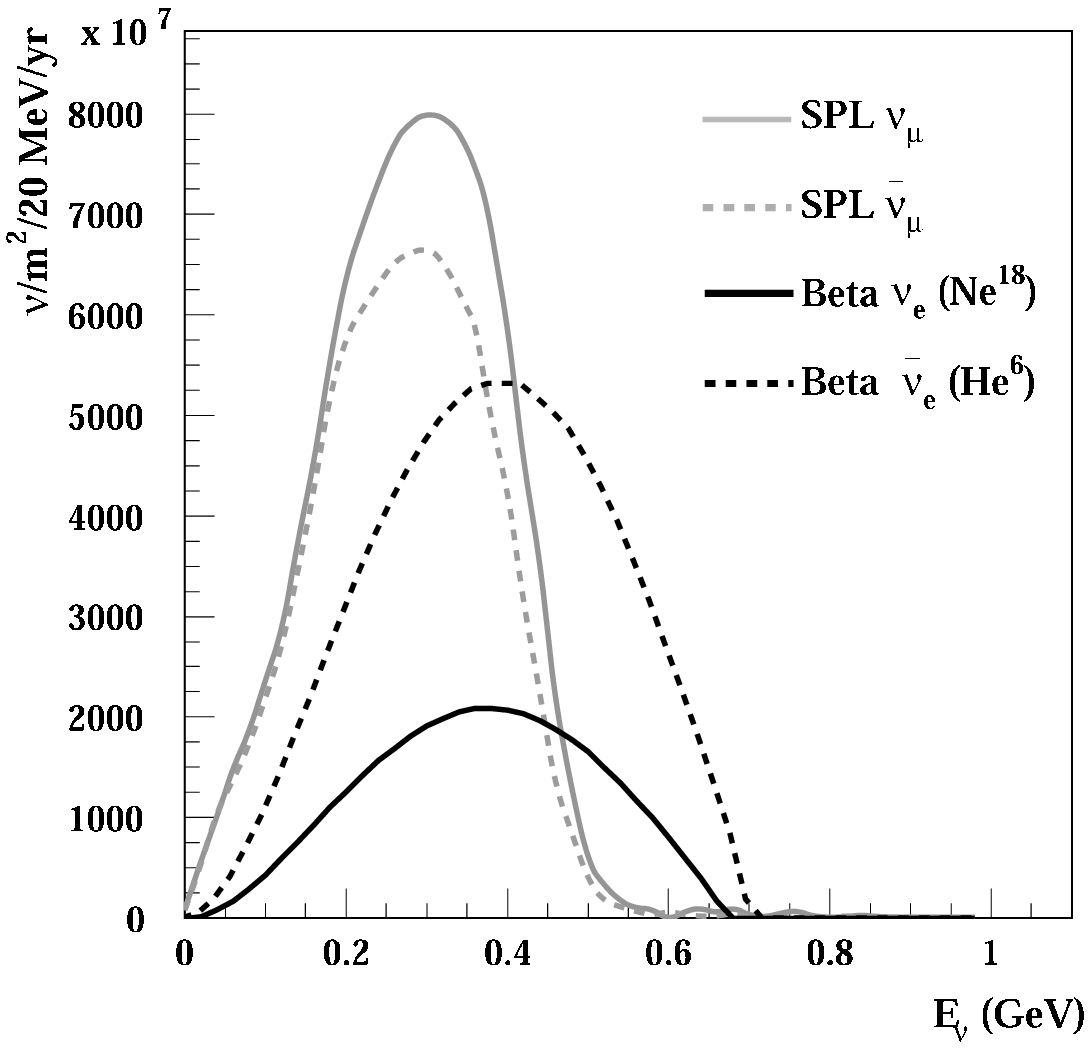}
  \mycaption{\label{fig:fluxComparison} 
  Comparison of the fluxes from SPL and \BB.}
\end{figure}

Considering the signal over square-root of background
ratio, the $3.5$~GeV beam energy is more favorable than the original
$2.2$~GeV option. Compared to the fluxes used in
Refs.~\cite{Mezzetto:2003mm,Donini:2004hu} the gain is at least a
factor $2.5$ and this justifies to reconsider in detail the physics
potential of the SPL Super Beam.
Both the appearance and the disappearance channels are used. For the
spectral analysis we use 10 bins of 100~MeV in the interval $0 < E_\nu
< 1$~GeV, applying the same migration matrices as for the \BB\ to take
into account properly the neutrino energy reconstruction. As ultimate
goal suggested in Ref.~\cite{T2K} a 2\% systematical error is used as
default both for signal and background, this would be achieved by a
special care of the design of the close position. However, we discuss
also how a 5\% systematical error affects the sensitivities.
Using neutrino cross-sections on water from Ref.~\cite{Nuance}, the
number of expected $\nu_\mu$ charged current is about $98$ per
kt~yr. In Fig.~\ref{fig:fluxComparison} we compare the fluxes from the
SPL to the one from the \BB.

\subsection{The atmospheric neutrino analysis}
\label{sec:atm-details}

The simulation of atmospheric neutrino data in MEMPHYS is based on the
analysis presented in Ref.~\cite{Huber:2005ep}, with the following
differences:
\begin{itemize}
  \item We replace the neutrino fluxes at Kamioka with those at Gran
    Sasso. We use the Honda calculations~\cite{Honda:2004yz}, which
    unfortunately are not yet available for the Fr\'ejus
    site. However, since the fluxes increase with the geomagnetic
    latitude and Fr\'ejus is northern than Gran Sasso, our choice is
    conservative.
    
  \item We take into account the specific geometry of the MEMPHYS
    detector. This is particularly important to properly separate
    fully contained from partially contained events, as well as
    stopping muon from through-going muon events.
    
  \item We divide our total data sample into 420 different bins:
    fully contained single-ring events, further subdivided according
    to flavor ($e$-like or $\mu$-like), lepton momentum (8 bins:
    0.1--0.3, 0.3--0.5, 0.5--1, 1--2, 2--3, 3--5, 5--8,
    8--$\infty$~GeV) and lepton direction (20 bins in zenith angle);
    fully contained multi-ring events, further subdivided according to
    flavor ($e$-like or $\mu$-like), reconstructed neutrino energy (3
    bins: 0--1.33, 1.33--5, 5--$\infty$~GeV) and lepton direction (10
    bins in zenith angle);
    partially contained $\mu$-like events, divided into 20 zenith bins;
    up-going muons, divided into stopping and through-going events, and in
    10 zenith bins each.

    \item We include in our calculations the neutral-current
      contamination of each bin. To this extent we assume that the
      ratio between neural-current and \emph{unoscillated}
      charged-current events in MEMPHYS is the same as in
      Super-Kamiokande, and we take this ratio from
      Ref.~\cite{Ashie:2005ik}.
      
    \item We consider also multi-ring events, which we define
      as fully contained charged-current events which are \emph{not}
      tagged as single-ring. Again, we assume that the survival
      efficiency and the NC contamination are the same as for
      Super-Kamiokande~\cite{Ashie:2005ik}.
\end{itemize}

The expected number of contained events is given by:
\begin{multline} \label{eq:contained}
    N_b(\vec\omega) = N_b^\text{NC} +
    n_\text{tgt} T \sum_{\alpha,\beta,\pm}
    \int_0^\infty dh \int_{-1}^{+1} dc_\nu
    \int_{\Emin}^\infty dE_\nu \int_{\Emin}^{E_\nu} dE_l
    \int_{-1}^{+1} dc_a \int_0^{2\pi} d\varphi_a
    \\
    \frac{d^3 \Phi_\alpha^\pm}{dE_\nu \, dc_\nu \, dh}(E_\nu, c_\nu, h)
    \, P_{\alpha\to\beta}^\pm(E_\nu, c_\nu, h \,|\, \vec\omega)
    \, \frac{d^2\sigma_\beta^\pm}{dE_l \, dc_a}(E_\nu, E_l, c_a)
    \, \varepsilon_\beta^b(E_l, c_l(c_\nu, c_a, \varphi_a))
    \,,
\end{multline}
where $P_{\alpha\to\beta}^+$ ($P_{\alpha\to\beta}^-$) is the
$\nu_\alpha \to \nu_\beta$ ($\bar{\nu}_\alpha \to \bar{\nu}_\beta$)
conversion probability for given values of the neutrino energy
$E_\nu$, the cosine $c_\nu$ of the angle between the incoming neutrino
and the vertical direction, the production altitude $h$, and the
neutrino oscillation parameters $\vec\omega$. We calculate the
conversion probability numerically in the general three-flavor
framework taking into account matter effects from a realistic Earth
density profile. Further, $N_b^\text{NC}$ is the neutral-current
background for the bin $b$, $n_\text{tgt}$ is the number of targets,
$T$ is the experiment running time, $\Phi_\alpha^+$ ($\Phi_\alpha^-$)
is the flux of atmospheric neutrinos (antineutrinos) of type $\alpha$,
and $\sigma_\beta^+$ ($\sigma_\beta^-$) is the charged-current
neutrino- (antineutrino-) nucleon interaction cross section.
The variable $E_l$ is the energy of the final lepton of type $\beta$,
while $c_a$ and $\varphi_a$ parametrize the opening angle between the
incoming neutrino and the final lepton directions as determined by the
kinematics of the neutrino interaction.
Finally, $\varepsilon_\beta^b$ gives the probability that a charged
lepton of type $\beta$, energy $E_l$ and direction $c_l$ contributes
to the bin $b$.

Up-going muon events are calculated as follows:
\begin{multline} \label{eq:upgoing}
    N_b(\vec\omega) = \rho_\text{rock} T \sum_{\alpha,\pm} 
    \int_0^\infty dh \int_{-1}^{+1} dc_\nu
    \int_{\Emin}^\infty dE_\nu 
    \int_{\Emin}^{E_\nu} dE^0_\mu \int_{\Emin}^{E^0_\mu} d\Efin_\mu
    \int_{-1}^{+1} dc_a \int_0^{2\pi} d\varphi_a
    \\
    \frac{d^3 \Phi_\alpha^\pm}{dE_\nu \, dc_\nu \, dh}(E_\nu, c_\nu, h)
    \, P_{\alpha\to\mu}^\pm(E_\nu, c_\nu, h \,|\, \vec\omega)
    \, \frac{d^2\sigma_\mu^\pm}{dE^0_\mu \, dc_a}(E_\nu, dE^0_\mu, c_a)
    \\
    \times R_\text{rock}(E^0_\mu,\Efin_\mu)
    \, \mathcal{A}_\text{eff}^b(\Efin_\mu,
    c_l(c_\nu, c_a, \varphi_a)) \,,
\end{multline}
where $\rho_\text{rock}$ is the density of targets in standard rock,
$R_\text{rock}$ is the effective muon range~\cite{Lipari:1991ut} for a
muon which is produced with energy $E^0_\mu$ and reaches the detector
with energy $\Efin_\mu$, and $\mathcal{A}_\text{eff}^b$ is the
effective area for the bin $b$. The other variables and physical
quantities are the same as for contained events.

The statistical analysis is based on the pull method, as described in
Ref.~\cite{Gonzalez-Garcia:2004wg}. In our analysis we include three
different kind of experimental uncertainties:
Flux uncertainties: total normalization (20\%), tilt factor (5\%),
zenith angle (5\%), $\nu/\bar\nu$ ratio (5\%), and $\mu/e$ ratio
(5\%);
cross-section uncertainties: total normalization (15\%) and $\mu/e$
ratio (1\%) for each type of charged-current interaction
(quasi-elastic, one-pion production, and deep-inelastic scattering),
and total normalization (15\%) for the neutral-current contributions;
systematic uncertainties: same as in previous analyses, details are
given in the Appendix of Ref.~\cite{Gonzalez-Garcia:2004wg}. In
addition, we assume independent normalization uncertainties (20\%) for
$e$-like and $\mu$-like multi-ring events.
Since we are dividing our data sample into a large number of bins, it
is important to use Poisson statistics as some of the bins contain
only a few number of events. We therefore write our $\chi^2$ as:
\begin{equation} \label{eq:poisson}
    \chi^2(\vec\omega) = \min_{\vec\xi} \left[ 2 \sum_b \left(
    N_b^\text{th}(\vec\omega,\, \vec\xi)
    - N_b^\text{ex} + N_b^\text{ex} \ln\frac{N_b^\text{ex}}
    {N_b^\text{th}(\vec\omega,\, \vec\xi)} \right) 
    + \sum_n \xi_n^2 \right] \,,
\end{equation}
where the number of events for a given value of the pulls $\vec\xi$ is
given by:
\begin{equation} \label{eq:theopulls}
    N_b^\text{th}(\vec\omega,\, \vec\xi) = N_b^\text{th}(\vec\omega) \,
    \exp\left( \sum_n \pi_b^n(\vec\omega)\, \xi_n \right) \,.
\end{equation}
The use of an exponential dependence on the pulls in
Eq.~\eqref{eq:theopulls}, rather than the usual linear dependence,
ensures that the theoretical predictions remain positive for
\emph{any} value of the pulls, thus avoiding numerical inconsistencies
during the pull minimization procedure.

\section{Degeneracies}
\label{sec:degeneracies}

\begin{figure}[!t]
\centering
\includegraphics[width=0.95\textwidth]{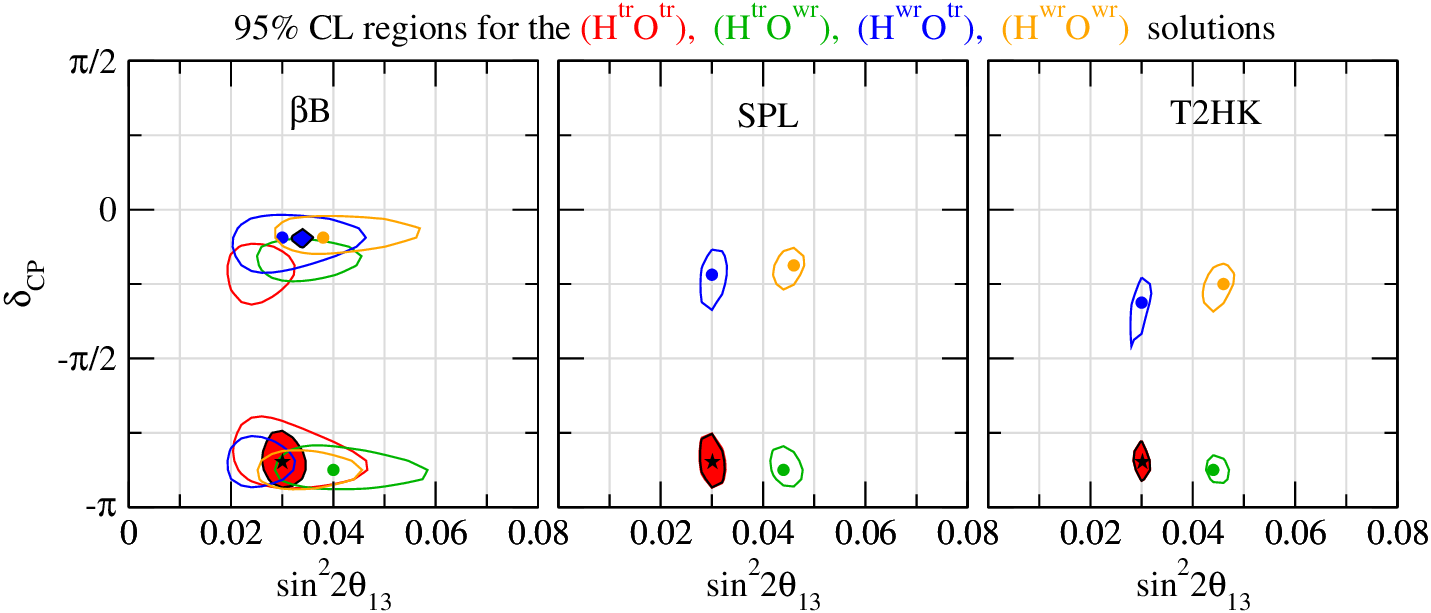}
  \mycaption{Allowed regions in $\sin^22\theta_{13}$ and
  $\delta_\mathrm{CP}$ for LBL data alone (contour lines) and LBL+ATM
  data combined (colored regions). $\mathrm{H^{tr/wr} (O^{tr/wr})}$
  refers to solutions with the true/wrong mass hierarchy (octant of
  $\theta_{23}$). The true parameter values are $\delta_\mathrm{CP} =
  -0.85 \pi$, $\sin^22\theta_{13} = 0.03$, $\sin^2\theta_{23} = 0.6$,
  and the values from Eq.~(\ref{eq:default-params}) for the other
  parameters. The running time is ($5\nu + 5\bar\nu$)~yrs for \BB\ and
  ($2\nu + 8\bar\nu$)~yrs for the Super Beams.}
\label{fig:degeneracies}
\end{figure}

A characteristic feature in the analysis of future LBL experiments is
the presence of {\it parameter degeneracies}.  Due to the inherent
three-flavor structure of the oscillation probabilities, for a given
experiment in general several disconnected regions in the
multi-dimensional space of oscillation parameters will be
present. Traditionally these degeneracies are referred to in the
following way:
\begin{itemize}
\item
The {\it intrinsic} or
($\delCP,\theta_{13}$)-degeneracy~\cite{Burguet-Castell:2001ez}:
For a measurement based on the $\nu_\mu \to \nu_e$ oscillation probability for
neutrinos and antineutrinos two disconnected solutions appear in the
($\delCP,\theta_{13}$) plane.
\item
The {\it hierarchy} or sign($\Delta
m^2_{31}$)-degeneracy~\cite{Minakata:2001qm}: The two solutions
corresponding to the two signs of $\Delta m^2_{31}$ appear in general
at different values of $\delCP$ and $\theta_{13}$.
\item
The {\it octant} or $\theta_{23}$-degeneracy~\cite{Fogli:1996pv}:
Since LBL experiments are sensitive mainly to $\sin^22\theta_{23}$ it
is difficult to distinguish the two octants $\theta_{23} < \pi/4$ and
$\theta_{23} > \pi/4$.  Again, the solutions corresponding to
$\theta_{23}$ and $\pi/2 - \theta_{23}$ appear in general at different
values of $\delCP$ and $\theta_{13}$.
\end{itemize}
This leads to an eight-fold ambiguity in $\theta_{13}$ and
$\delCP$~\cite{Barger:2001yr}, and hence degeneracies provide a
serious limitation for the determination of $\theta_{13}$, $\delCP$,
and the sign of $\Delta m^2_{31}$. Recent discussions of degeneracies
can be found for example in
Refs.~\cite{Huber:2002mx,Huber:2005ep,Yasuda:2004gu,Ishitsuka:2005qi};
degeneracies in the context of CERN--Fr\'ejus \BB\ and SPL have been
considered previously in Ref.~\cite{Donini:2004hu}.
In Fig.~\ref{fig:degeneracies} we illustrate the effect of
degeneracies for the \BB, SPL, and T2HK experiments. Assuming the
true parameter values $\delta_\mathrm{CP} = -0.85 \pi$,
$\sin^22\theta_{13} = 0.03$, $\sin^2\theta_{23} = 0.6$ we show the
allowed regions in the plane of $\stheta$ and $\delCP$ taking into
account the solutions with the wrong hierarchy and the wrong octant of
$\theta_{23}$.

\begin{figure}[!t]
\centering
\includegraphics[width=0.9\textwidth]{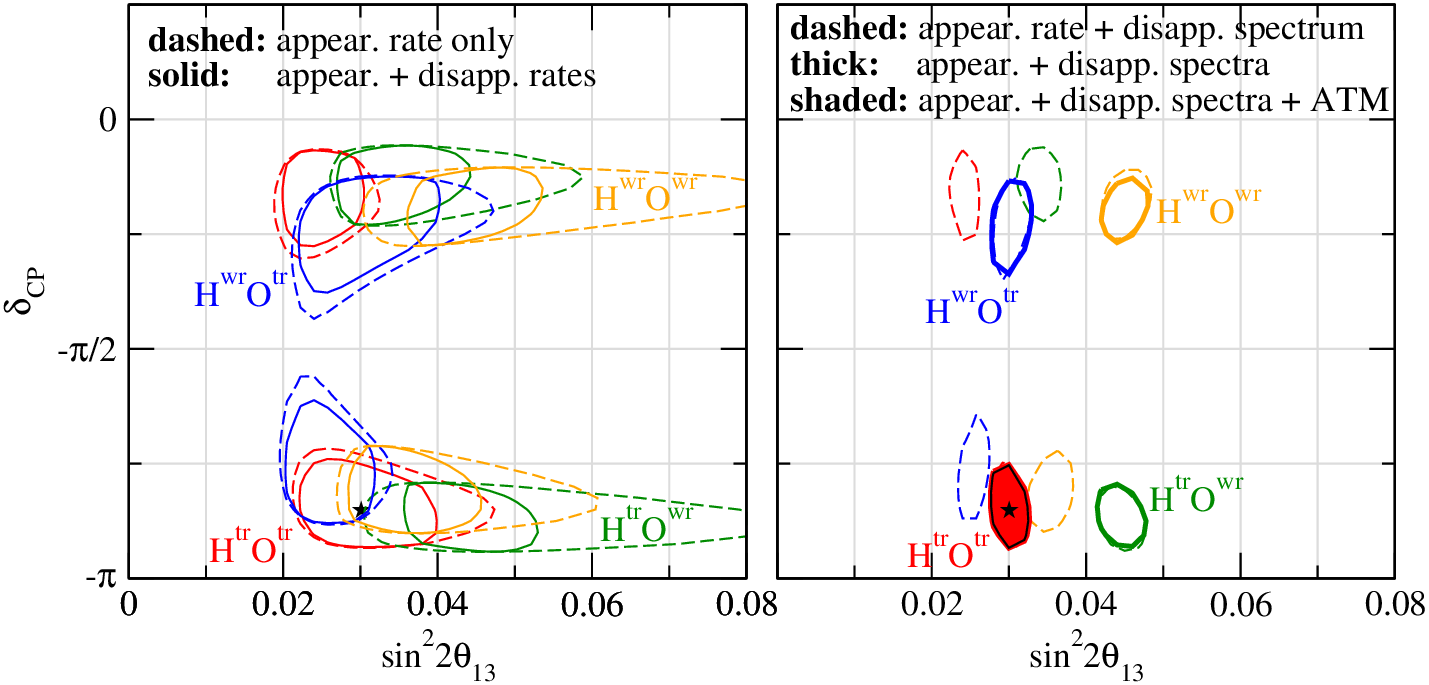}
  \mycaption{Resolving degeneracies in SPL by successively using the
  appearance rate measurement, disappearance channel rate and
  spectrum, spectral information in the appearance channel, and
  atmospheric neutrinos.  Allowed regions in $\sin^22\theta_{13}$ and
  $\delta_\mathrm{CP}$ are shown at 95\%~CL, and $\mathrm{H^{tr/wr}
  (O^{tr/wr})}$ refers to solutions with the true/wrong mass hierarchy
  (octant of $\theta_{23}$). The true parameter values are
  $\delta_\mathrm{CP} = -0.85 \pi$, $\sin^22\theta_{13} = 0.03$,
  $\sin^2\theta_{23} = 0.6$, and the values from
  Eq.~(\ref{eq:default-params}) for the other parameters. The running
  time is ($2\nu + 8\bar\nu$)~yrs.

}
\label{fig:degeneracies_SPL}
\end{figure}

As visible in Fig.~\ref{fig:degeneracies} for the
Super Beam experiments SPL and T2HK there is only a four-fold
degeneracy related to sign($\Delta m^2_{31}$) and the octant of
$\theta_{23}$, whereas the intrinsic degeneracy can be resolved. 
Several pieces of information contribute to this effect, as we
illustrate at the example of SPL in Fig.~\ref{fig:degeneracies_SPL}.
The dashed curves in the left panel of this figure show the allowed
regions for only the appearance measurement (for neutrinos and
antineutrinos) without spectral information, i.e., just a counting
experiment. In this case the eight-fold degeneracy is present in its
full beauty, and one finds two solutions (corresponding to the
intrinsic degeneracy) for each choice of sign($\Delta m^2_{31}$) and
the octant of $\theta_{23}$. Moreover, the allowed regions are
relatively large. For the thin solid curves the information from the
disappearance rate is added. The main effect is to decrease the size
of the allowed regions in $\stheta$. This is especially pronounced for
the solutions involving the wrong octant of $\theta_{23}$, since these
solutions are strongly affected by an uncertainty in $\theta_{23}$
which gets reduced by the disappearance information. Using in addition
to the disappearance rate also the spectrum again decreases the size
of the allowed regions, however, still all eight solutions are present
(compare dashed curves in the right panel).
The most relevant effect comes from the inclusion of spectral
information in the appearance channel, as visible from the comparison
of the dashed and thick-solid curves in the right panel of
Fig.~\ref{fig:degeneracies_SPL}. The intrinsic degeneracy gets
resolved and only four solutions corresponding to the sign and octant
degeneracies are left (see, e.g.,
Refs.~\cite{Huber:2002mx,Huber:2005ep,Ishitsuka:2005qi}).\footnote{The
inclusion of spectral information might be the source of possible
differences to previous studies, see e.g.\ Ref.~\cite{Donini:2004hu}.}
Note that the thick curves in the right panel of
Fig.~\ref{fig:degeneracies_SPL} correspond to the regions show in
Fig.~\ref{fig:degeneracies} for the SPL.
Finally, by the inclusion of information from atmospheric neutrinos
all degeneracies can be resolved in this example, and the true
solution is identified at 95\%~CL (see Sec.~\ref{sec:atmospherics} and
Ref.~\cite{Huber:2005ep} for further discussions of atmospheric
neutrinos).

Concerning the \BB\ one observes from Fig.~\ref{fig:degeneracies} that
in this case the ($\delCP,\theta_{13}$)-degeneracy cannot be resolved
and one has to deal with eight distinct solutions. One reason for this
is the absence of precise information on $|\Delta m^2_{31}|$ and
$\sin^22\theta_{23}$ which is provided by the $\nu_\mu$ disappearance
in Super Beam experiments but is not available from the \BB. If
external information on these parameters at the level of 3\% is
included the allowed regions in Fig.~\ref{fig:degeneracies} are
significantly reduced. However, still all eight solutions are present,
which indicates that for the \BB\ spectral information is not
efficient enough to resolve the ($\delCP,\theta_{13}$)-degeneracy, and
in this case only the inclusion of atmospheric neutrino data allows a
nearly complete resolution of the degeneracies.

An important observation from Fig.~\ref{fig:degeneracies} is that
degeneracies have only a very small impact on the CP violation
discovery, in the sense that if the true solution is CP violating also
the fake solutions are located at CP violating values of
$\delCP$. Indeed, since for the relatively short baselines in the
experiments under consideration matter effects are very small, the
sign($\Delta m^2_{31}$)-degenerate solution is located within good
approximation at $\delCP' \approx \pi -
\delCP$~\cite{Minakata:2001qm}. Therefore, although degeneracies
strongly affect the determination of $\theta_{13}$ and $\delCP$ they
have only a small impact on the CP violation discovery potential.
Furthermore, as clear from Fig.~\ref{fig:degeneracies} the sign($\Delta
m^2_{31}$) degeneracy has practically no effect on the $\theta_{13}$
measurement, whereas the octant degeneracy has very little impact on
the determination of $\delCP$.

\begin{figure}[!t]
\centering
\includegraphics[width=0.9\textwidth]{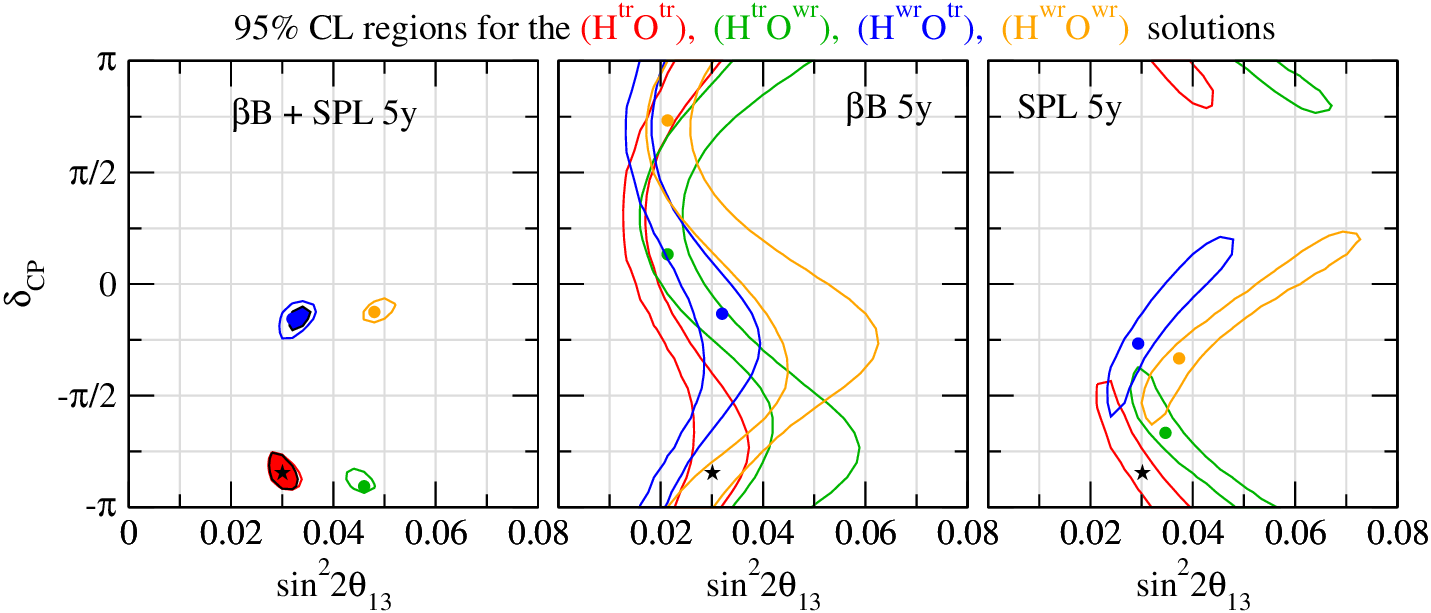}
  \mycaption{Allowed regions in $\sin^22\theta_{13}$ and
  $\delta_\mathrm{CP}$ for 5~years data (neutrinos only) from \BB,
  SPL, and the combination. $\mathrm{H^{tr/wr} (O^{tr/wr})}$ refers to
  solutions with the true/wrong mass hierarchy (octant of
  $\theta_{23}$). For the colored regions in the left panel also
  5~years of atmospheric data are included; the solution with the
  wrong hierarchy has $\Delta\chi^2 = 4$. The true parameter
  values are $\delta_\mathrm{CP} = -0.85 \pi$, $\sin^22\theta_{13} =
  0.03$, $\sin^2\theta_{23} = 0.6$, and the values from
  Eq.~(\ref{eq:default-params}) for the other parameters. For the \BB\
  only analysis (middle panel) an external accuracy of 2\% (3\%) for
  $|\Delta m^2_{31}|$ ($\theta_{23}$) has been assumed, whereas for
  the left and right panel the default value of 10\% has been used.}
\label{fig:degeneracies_5yrs}
\end{figure}

Fig.~\ref{fig:degeneracies} shows also that the fake solutions occur
at similar locations in the ($\stheta$, $\delCP$) plane for \BB\ and
SPL. Therefore, as noted in Ref.~\cite{Donini:2004hu}, in this sense
the two experiments are not complementary, and the combination of
10~years of \BB\ and SPL data is not very effective in resolving
degeneracies. This is obvious since the baseline is the same and the
neutrino energies are similar.
Note however, that the \BB\ looks for $\nu_e\to\nu_\mu$ appearance,
whereas in SPL the T-conjugate channel $\nu_\mu\to\nu_e$ is observed.
Assuming CPT invariance the relation $P_{\nu_\alpha\to\nu_\beta} =
P_{\bar\nu_\beta\to\bar\nu_\alpha}$ holds, which implies that the
antineutrino measurement can be replaced by a measurement in the
T-conjugate channel.  Hence, if \BB\ and SPL experiments are available
simultaneously the full information can be obtained just from neutrino
data, and in principle the (time consuming) antineutrino measurement 
is not necessary. As shown in Fig.~\ref{fig:degeneracies_5yrs} the
combination of 5~yrs neutrino data from the \BB\ with 5~yrs of
neutrino data from SPL leads to a result very close to the 10~yrs
neutrino+antineutrino data from one experiment alone. Hence, if \BB\
and SPL experiments are available simultaneously the data taking
period is reduced approximately by a factor of 2 with respect to a
single experiment. This synergy is discussed later in
Sec.~\ref{sec:synergies-beams} in the context of the $\theta_{13}$ and
CP violation discovery potentials.

\section{Physics potential}
\label{sec:sensitivities}

\subsection{Sensitivity to the atmospheric parameters}
\label{sec:atm}

\begin{figure}[!t]
\centering
  \includegraphics[width=0.55\textwidth]{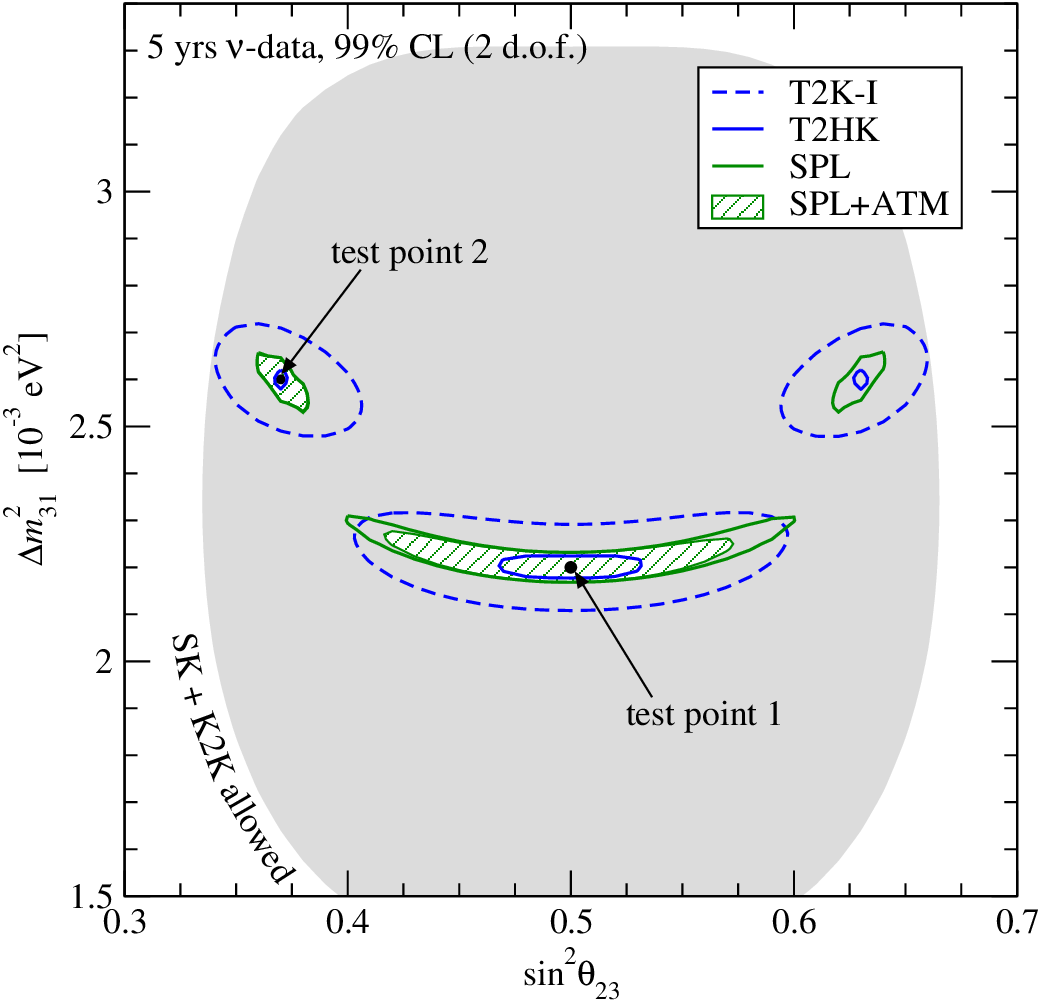}
  \mycaption{\label{fig:atm-params} Allowed regions of $\Delta
  m^2_{31}$ and $\sin^2\theta_{23}$ at 99\%~CL (2 d.o.f.)  after 5~yrs
  of neutrino data taking for SPL, T2K phase~I, T2HK, and the
  combination of SPL with 5~yrs of atmospheric neutrino data in the
  MEMPHYS detector. For the true parameter values we use $\Delta
  m^2_{31} = 2.2\, (2.6) \times 10^{-3}~\mathrm{eV}^2$ and
  $\sin^2\theta_{23} = 0.5 \, (0.37)$ for the test point 1 (2), and
  $\theta_{13} = 0$ and the solar parameters as given in
  Eq.~(\ref{eq:default-params}). The shaded region corresponds to the
  99\%~CL region from present SK and K2K data~\cite{Maltoni:2004ei}.}
\end{figure}

The $\nu_\mu$ disappearance channel available in the Super Beam
experiments SPL and T2HK allows a precise determination of the
atmospheric parameters $|\Delta m^2_{31}|$ and $\sin^22\theta_{23}$,
see, e.g., Refs.~\cite{Antusch:2004yx,Minakata:2004pg,Donini:2005db}
for recent analyses). Fig.~\ref{fig:atm-params} illustrates the
improvement on these parameters by Super Beam experiments with respect
to the present knowledge from SK atmospheric and K2K data. We show the
allowed regions at 99\%~CL for T2K-I, SPL, and T2HK, where in all
three cases 5~years of neutrino data are assumed. T2K-I corresponds to
the phase~I of the T2K experiment with a beam power of 0.77~MW and the
Super-Kamiokande detector as target~\cite{T2K}. In Tab.~\ref{tab:atm-params} we
give the corresponding relative accuracies at 3$\sigma$ for $|\Delta
m^2_{31}|$ and $\sin^2\theta_{23}$.

\begin{table}[!t]
  \centering
  \begin{tabular}{lcrrr}
  \hline\noalign{\smallskip}
    & True values  & T2K-I & SPL & T2HK \\
  \noalign{\smallskip}\hline\noalign{\smallskip}
  $\Delta m^2_{31}$   & $2.2\cdot 10^{-3}$ eV$^2$ & 4.7\% & 3.2\% & 1.1\% \\
  $\sin^2\theta_{23}$ & $0.5$                     & 20\%  & 20\%  & 6\%   \\
  \noalign{\smallskip}\hline\noalign{\smallskip}
  $\Delta m^2_{31}$   & $2.6\cdot 10^{-3}$ eV$^2$ & 4.4\% & 2.5\% & 0.7\% \\
  $\sin^2\theta_{23}$ & $0.37$                    & 8.9\% & 3.1\% & 0.8\% \\
  \noalign{\smallskip}\hline
  \end{tabular}
  \mycaption{Accuracies at $3\sigma$ on the atmospheric parameters
  $|\Delta m^2_{31}|$ and $\sin^2\theta_{23}$ for 5 years of neutrino
  data from T2K-I, SPL, and T2HK for the two test points shown in
  Fig.~\ref{fig:atm-params} ($\theta^\mathrm{true}_{13} = 0$). The
  accuracy for a parameter $x$ is defined as $(x^\mathrm{upper} -
  x^\mathrm{lower})/(2 x^\mathrm{true})$, where $x^\mathrm{upper}$
  ($x^\mathrm{lower}$) is the upper (lower) bound at 3$\sigma$ for
  1~d.o.f.\ obtained by projecting the contour $\Delta \chi^2 = 9$
  onto the $x$-axis. For the accuracies for test point~2 the octant
  degenerate solution is neglected.\label{tab:atm-params}}
\end{table}

From the figure and the table it becomes evident that the T2K setups
are very good in measuring the atmospheric parameters, and only a
modest improvement is possible with SPL with respect to T2K phase~I.
T2HK provides an excellent sensitivity for these parameters, and for
the example of the test point~2 sub-percent accuracies are obtained at
3$\sigma$. The disadvantage of SPL with respect to T2HK is the
limited spectral information. Because of the lower beam energy
nuclear Fermi motion is a severe limitation for energy reconstruction
in SPL, whereas in T2K the somewhat higher energy allows an efficient
use of spectral information of quasi-elastic events. Indeed, due to
the large number of events in the disappearance channel (cf.\
Tab.~\ref{tab:events}) the measurement is completely dominated by the
spectrum, and even increasing the normalization uncertainty up to
100\% has very little impact on the allowed regions. The effect of
spectral information on the disappearance measurement is
discussed in some detail in Ref.~\cite{Donini:2005db}.

In the interpretation of the numbers given in
Tab.~\ref{tab:atm-params} one should consider that at accuracies below
1\% systematics might become important, which are not accounted for
here. We do include the most relevant systematics (see
Secs.~\ref{sec:analysis} and \ref{sec:experiments}), however, at that
level additional uncertainties related to, for example, the spectral
shapes of signal and/or background, or the energy calibration might
eventually limit the accuracy.

For the test point~1, with maximal mixing for $\theta_{23}$, rather
poor accuracies of $\sim20\%$ for T2K-I and SPL, and $6\%$ for T2HK
are obtained for $\sin^2\theta_{23}$. The reason is that in the
disappearance channel $\sin^22\theta_{23}$ is measured with high
precision, which translates to rather large errors for
$\sin^2\theta_{23}$ if $\theta_{23} = \pi/4$~\cite{Minakata:2004pg}.
For the same reason it is difficult to resolve the octant degeneracy,
and for the test point~2, with a non-maximal value of
$\sin^2\theta_{23} = 0.37$, for all three LBL experiments the
degenerate solution is present around $\sin^2\theta_{23} = 0.63$.
As pointed out in Refs.~\cite{Peres:2003wd,Gonzalez-Garcia:2004cu}
atmospheric neutrino data may allow to distinguish between the two
octants of $\theta_{23}$. If 5~years of atmospheric neutrino data in
MEMPHYS are added to the SPL data, the degenerate solution for the
test point~2 can be excluded at more than $5\sigma$ and hence the
octant degeneracy is resolved in this example, see
Sec.~\ref{sec:atmospherics} for a more detailed discussion.

\subsection{The $\theta_{13}$ discovery potential}
\label{sec:th13}

If no finite value of $\theta_{13}$ is discovered by the next round of
experiments an important task of the experiments under consideration
here is to push further the sensitivity to this parameter. In this
section we address this problem, where we use to following definition
of the $\theta_{13}$ discovery potential: Data are simulated for a
finite true value of $\stheta$ and a given true value for $\delCP$. If
the $\Delta\chi^2$ of the fit to these data with $\theta_{13} = 0$ is
larger than 9 the corresponding true value of $\theta_{13}$ ``is
discovered at 3$\sigma$''. In other words, the $3\sigma$ discovery
limit as a function of the true $\delCP$ is given by the true value of
$\stheta$ for which $\Delta\chi^2(\theta_{13}=0) = 9$. In the fitting
process we minimize the $\Delta\chi^2$ with respect to $\theta_{12}$,
$\theta_{23}$, $\Delta m^2_{12}$, and $\Delta m^2_{31}$, and in
general one has to test also for degenerate solutions in sign($\Delta
m^2_{31}$) and the octant of $\theta_{23}$.

\begin{figure}
  \centering \includegraphics[width=0.9\textwidth]{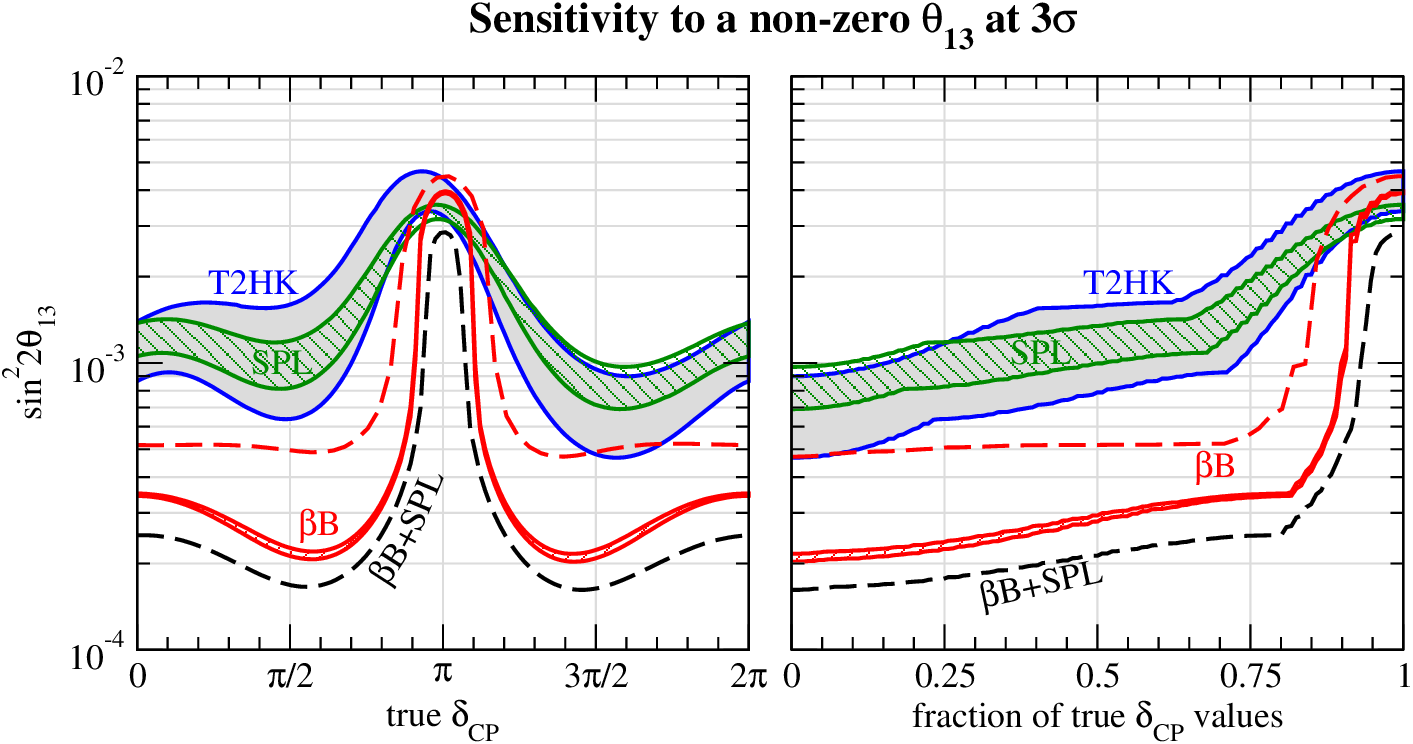}
  \mycaption{$3\sigma$ discovery sensitivity to $\stheta$ for \BB,
  SPL, and T2HK as a function of the true value of \delCP\ (left
  panel) and as a function of the fraction of all possible values of
  \delCP\ (right panel). The running time is ($5\nu + 5\bar\nu$)~yrs
  for \BB\ and ($2\nu + 8\bar\nu$)~yrs for the Super Beams. The width
  of the bands corresponds to values for the systematical errors
  between 2\% and 5\%. The black curves correspond to the combination
  of \BB\ and SPL with 10~yrs of total data taking each for a
  systematical error of 2\%, and the dashed curves show the
  sensitivity of the \BB\ when the number of ion decays/yr are reduced
  by a factor of two with respect to the values given in
  Tab.~\ref{tab:setups}.\label{fig:th13}}
\end{figure}

The discovery limits are shown for \BB, SPL, and T2HK in
Fig.~\ref{fig:th13}. One observes that SPL and T2HK are rather similar
in performance, whereas the \BB\ with our standard fluxes performs
significantly better. For all three facilities a guaranteed discovery
reach of $\stheta \simeq 5\times 10^{-3}$ is obtained, irrespective of
the actual value of \delCP, however, for certain values of \delCP\ the
sensitivity is significantly improved. For SPL and T2HK discovery
limits around $\stheta \simeq 10^{-3}$ are possible for a large
fraction of all possible values of \delCP, whereas for our standard
\BB\ a sensitivity below $\stheta = 4\times 10^{-4}$ is reached for
80\% of all possible values of \delCP. If 10~years of data from \BB\
and SPL are combined the discovery limit is dominated by the \BB.
Let us stress that the \BB\ performance depends crucially on the
neutrino flux intensity, as can be seen from the dashed curves in
Fig.~\ref{fig:th13}, which has been obtained by reducing the number of
ion decays/yr by a factor of two with respect to our standard values
given in Tab.~\ref{tab:setups}. In this case the sensitivity decreases
significantly, but still values slightly better than from the
Super Beam experiments are reached.

The peak of the sensitivity curves around $\delCP \approx \pi$ appears
due to the interplay of neutrino and antineutrino data. For the Super
Beams neutrino (antineutrino) data are most sensitive in the region
$\pi \lesssim \delCP \lesssim 2\pi$ ($0 \lesssim \delCP \lesssim
\pi$), and opposite for the \BB, compare also Fig.~\ref{fig:th13-5yrs}
in Sec.~\ref{sec:synergies-beams}. The particular shape of the
sensitivity curves emerges from the relative location of the
corresponding curves for neutrino and antineutrino data, which is
controlled by the $L/E_\nu$ value where the experiment is operated and
the value of $|\Delta m^2_{31}|$. The fact that the peak is most
pronounced for the \BB\ follows from the somewhat smaler $L/E_\nu$ of
the \BB\ compared to the Super Beams, whereas the shapes for SPL and
T2HK are similar because of the similar $L/E_\nu$ values.

In Fig.~\ref{fig:th13} we illustrate also the effect of systematical
errors on the $\theta_{13}$ discovery reach. The lower boundary of the
band for each experiment corresponds to a systematical error of 2\%,
whereas the upper boundary is obtained for 5\%. These errors include
the (uncorrelated) normalization uncertainties on the signal as well
as on the background, where the crucial uncertainty is the error on
the background. We find that the \BB\ is basically not affected by
these errors, since the background has a rather different spectral
shape (strongly peaked at low energies) than the signal. The fact
that T2HK is relatively strongly affected by the actual value of the
systematics can by understood by considering the ratio of signal to
the square-root of the background using the numbers of
Tab.~\ref{tab:events}. We shall discuss this issue in more detail in
the next section in the context of the CP violation discovery reach.

\begin{figure}
  \centering \includegraphics[width=0.9\textwidth]{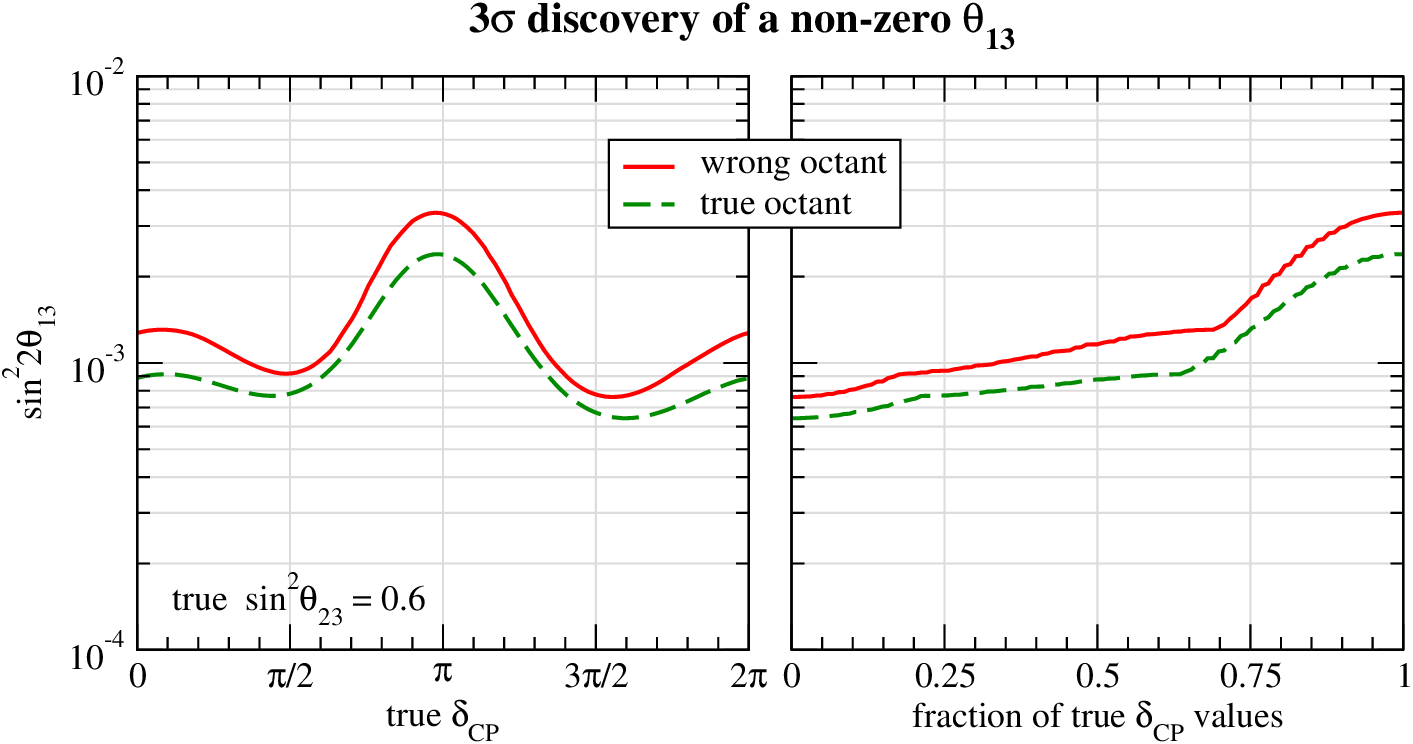}
  \mycaption{\label{fig:SPLTheta13Disco} $3\sigma$ discovery
  sensitivity to $\stheta$ for the SPL as a function of the true value
  of \delCP\ for $\sin^2\theta_{23}^\mathrm{true} = 0.6$ and true
  values for the other parameters as given in
  Eq.~(\ref{eq:default-params}). The running time is ($2\nu +
  8\bar\nu$)~yrs.
}
\end{figure}

Let us remark that the $\theta_{13}$ sensitivities are practically not
affected by the sign($\Delta m^2_{31}$)-degeneracy. This is easy to
understand, since the data is fitted with $\theta_{13} = 0$, and in
this case both mass hierarchies lead to very similar event rates. If
the inverted hierarchy is used as the true hierarchy, the peak in the
discovery limit visible in the left panel of Fig.~\ref{fig:th13}
around $\delCP \sim \pi$ moves to $\delCP \sim 0$. However, the
characteristic shape of the curves, and in particular, the sensitivity
as a function of the \delCP-fraction shown in the right panel are
hardly affected by the sign of the true $\Delta m^2_{31}$.
In case of a non-maximal value of $\theta_{23}$ the octant-degeneracy
has a minor impact on the $\theta_{13}$ discovery potential, as
illustrated in Fig.~\ref{fig:SPLTheta13Disco} for the SPL. We show the
discovery limit obtained with the true and the fake octant of
$\theta_{23}$ for a true value of $\sin^2\theta_{23}= 0.6$. Let us
note that for true values of $\sin^2\theta_{23} > 0.5$ the
octant-degenerate solution leads to a worse sensitivity to
$\theta_{13}$ (see figure), whereas for $\sin^2\theta_{23} < 0.5$ the
fake solution does not affect the $\theta_{13}$ discovery, since in
this case the sensitivity is increased.

\subsection{Sensitivity to CP violation}
\label{sec:CPV}

In case a finite value of $\theta_{13}$ is established it is important
to quantitatively assess the discovery potential for leptonic CP
violation (CPV). The CP symmetry is violated if the complex phase
\delCP\ is different from $0$ and $\pi$. Therefore, CPV is discovered
if these values for \delCP\ can be excluded. 
We evaluate the discovery potential for CPV in the following way:
Data are calculated by scanning the true values of $\stheta$ and
$\delCP$. Then these data are fitted with the CP conserving values
$\delCP = 0$ and $\delCP = \pi$, where all parameters except \delCP\
are varied and the sign and octant degeneracies are taken into
account. If no fit with $\Delta \chi^2 < 9$ is found CP conserving
values of \delCP\ can be excluded at $3\sigma$ for the chosen values
of $\delta_\mathrm{CP}^\mathrm{true}$ and $\stheta^\mathrm{true}$.

\begin{figure}[!t]
  \centering
   \includegraphics[width=0.65\textwidth]{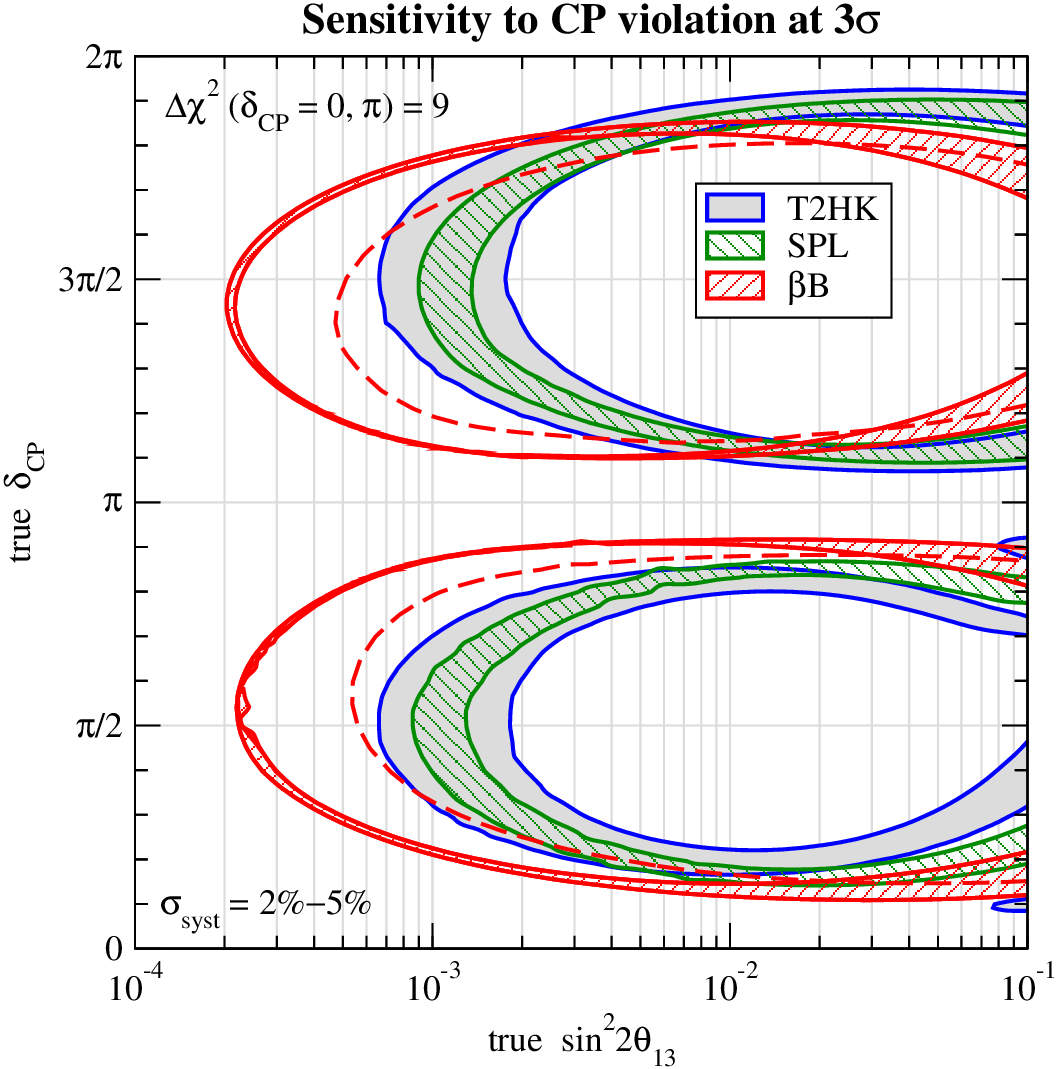}
   \mycaption{CPV discovery potential for \BB, SPL, and T2HK: For
   parameter values inside the ellipse-shaped curves CP conserving
   values of \delCP\ can be excluded at $3\sigma$ $(\Delta\chi^2>9)$.
   The running time is ($5\nu + 5\bar\nu$)~yrs for \BB\ and ($2\nu +
   8\bar\nu$)~yrs for the Super Beams. The width of the bands
   corresponds to values for the systematical errors from 2\% to
   5\%. The dashed curves show the sensitivity of the \BB\ when the
   number of ion decays/yr are reduced by a factor of two with respect
   to the values given in Tab.~\ref{tab:setups} for 2\%
   systematics.\label{fig:CPV}}
\end{figure}

The CPV discovery potential for \BB, SPL, and T2HK is shown in
Fig.~\ref{fig:CPV}. As in the case of the $\theta_{13}$ sensitivity we
find that SPL and T2HK perform rather similar, whereas the \BB\ has
significantly better sensitivity if our adopted numbers of ion decays
per year can be achieved. For systematical errors of 2\% maximal CPV
(for $\delCP^\mathrm{true} = \pi/2, \, 3\pi/2$) can be discovered at
$3\sigma$ down to $\stheta \simeq 8.8 \,(6.6)\times 10^{-4}$ for SPL
(T2HK), and $\stheta \simeq 2\times 10^{-4}$ for the \BB. This number
for the \BB\ is increased by a factor 3 if the fluxes are reduced to
half of our nominal values.  The best sensitivity to CPV is obtained
for all three facilities around $\stheta \sim 10^{-2}$. For this value
CPV can be established for 78\%, 73\%, 75\% of all values of \delCP\
for \BB, SPL, T2HK, respectively (again for systematics of 2\%).

The widths of the bands in Fig.~\ref{fig:CPV} corresponds to different
values for systematical errors. The curves which give the best
sensitivities are obtained for systematics of 2\%, the curves
corresponding to the worst sensitivity have been computed for
systematics of 5\%. We change the uncertainty on the signal as well as
on the background, however, it turns out that the most relevant
uncertainty is the background normalization. We find that the impact
of systematics is very small for the \BB. The reason for this is that
the spectral shape of the background in the \BB\ (from pions and
atmospheric neutrinos) is very different from the signal, and
therefore they can be disentangled by the fit of the energy spectrum.
For the Super Beams the background spectrum is more similar to the
signal, and therefore an uncertainty on the background normalization
might have a strong impact on the sensitivity, as visible from the SPL
and T2HK curves in Fig.~\ref{fig:CPV}. In particular T2HK is strongly
affected, and moving from 2\% to 5\% uncertainy decreases the
sensitivity to maximal CPV by a factor 3.

\begin{figure}[!t]
  \centering
   \includegraphics[width=0.9\textwidth]{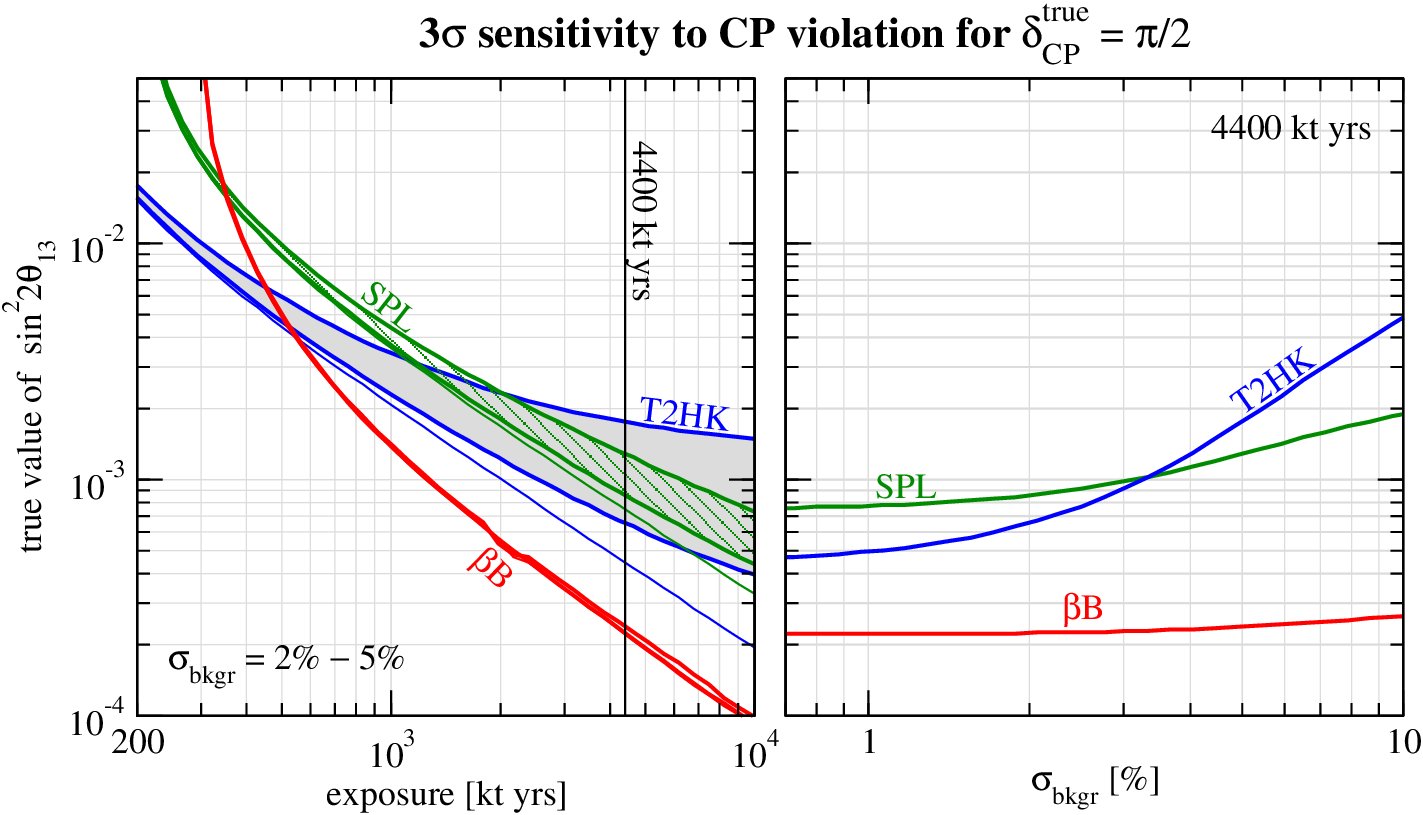}
   \mycaption{Impact of total exposure and systematical errors on the
   CPV discovery potential of \BB, SPL, and T2HK. We show the
   smallest true value of $\stheta$ for which $\delCP = \pi/2$ can be
   distinguished from $\delCP = 0$ or $\delCP = \pi$ at $3\sigma$
   $(\Delta\chi^2>9)$ as a function of the exposure in kt~yrs (left)
   and as a function of the systematical error on the background
   $\sigma_\mathrm{bkgr}$ (right). The widths of the curves in the
   left panel corresponds to values of $\sigma_\mathrm{bkgr}$ from 2\%
   to 5\%. The thin solid curves in the left panel corresponds to no
   systematical errors. The right plot is calculated for the standard
   exposure of 4400~kt~yrs. No systematical error on the signal has
   been assumed. \label{fig:systematics}}
\end{figure}

This interesting feature can be understood in the following way. A
rough measure to estimate the sensitivity is given by the signal
compared to the error on the background. The latter receives
contributions from the statistical error $\sqrt{B}$ and from the
systematical uncertainty $\sigma_\mathrm{bkgr}B$, where $B$ is the
number of background events and $\sigma_\mathrm{bkgr}$ is the
(relative) systematical error. Hence the importance of the systematics
can be estimated by the ratio of systematical and statistical errors
$\sigma_\mathrm{bkgr} B / \sqrt{B} = \sigma_\mathrm{bkgr} \sqrt{B}$.
Summing the numbers for background events in the neutrino and
antineutrino channels given in Tab.~\ref{tab:events} one finds that
systematical errors dominate ($\sigma_\mathrm{bkgr} \sqrt{B} > 1$) if
$\sigma_\mathrm{bkgr} \gtrsim 6\%,\, 3\%, \, 2\%$ for \BB, SPL, T2HK,
respectively. 
In the right panel Fig.~\ref{fig:systematics} we show the sensitivity
to maximal CPV (as defined in the figure caption) as a function of
$\sigma_\mathrm{bkgr}$. Indeed, the worsening of the sensitivity due
to systematics occurs roughly at the values of $\sigma_\mathrm{bkgr}$
as estimated above. For a more quantitative understanding of these
curves it is necessary to consider the number of signal and background
events for neutrinos and antineutrinos separately, as well as to take
into account spectral information.

The left panel of Fig.~\ref{fig:systematics} shows the sensitivity to
maximal CPV as a function of the exposure\footnote{Note that the CPV
sensitivity for the \BB\ with reduced fluxes from Fig.~\ref{fig:CPV}
is worse than the value which follows from Fig.~\ref{fig:systematics}.
The reason is that in Fig.~\ref{fig:systematics} the total exposure is
scaled (mass~$\times$~time), i.e., signal and background are scaled in
the same way, whereas for the dashed curve in Fig.~\ref{fig:CPV} only
the fluxes are reduced but backgrounds are kept constant.} for values
of $\sigma_\mathrm{bkgr}$ from 2\% to 5\%. One can observe clearly
that for the standard exposure of 4400~kt~yrs T2HK is dominated by
systematics and changing $\sigma_\mathrm{bkgr}$ from 2\% to 5\% has a
big impact on the sensitivity. In contrast the CERN--MEMPHYS
experiments (especially the \BB) are rather stable with respect to
systematics and for the standard exposure they are still statistics
dominated. We conclude that in T2HK systematics have to be under very
good control, whereas this issue is less important for \BB\ and SPL.
We have checked explicitly that the systematical error on the signal
has negligible impact on these results. Therefore, we have set this
error to zero for calculating Fig.~\ref{fig:systematics} to highlight
the importance of the background error. In all other calculations also
the signal error is included, in particular also in Fig.~\ref{fig:CPV}.

Let us remark that for the T2KK configuration (with one half of the
Hyper-K detector mass at Kamioka and the second half at the same
off-axis angle in Korea) the problem of systematics might be less
severe than for T2HK, since both detectors observe the same flux and
background. 
Note however, that an important issue for the CPV sensitivity is
whether systematics between neutrino and antineutrino data are
correlated or not. We have checked that the worse sensitivies for T2HK
shown in Fig.~\ref{fig:CPV} compared to the results obtained in
Refs.~\cite{Ishitsuka:2005qi,Kajita:2006bt} can be traced back to the
fact that in these papers neutrino and antineutrino systematics are
correlated, whereas we consider them to be independent. Note that our
approach is conservative, and the assumption of uncorrelated errors
has been adopted also for the CERN--MEMPHYS experiments.

\begin{figure}[!t]
  \centering
   \includegraphics[width=0.8\textwidth]{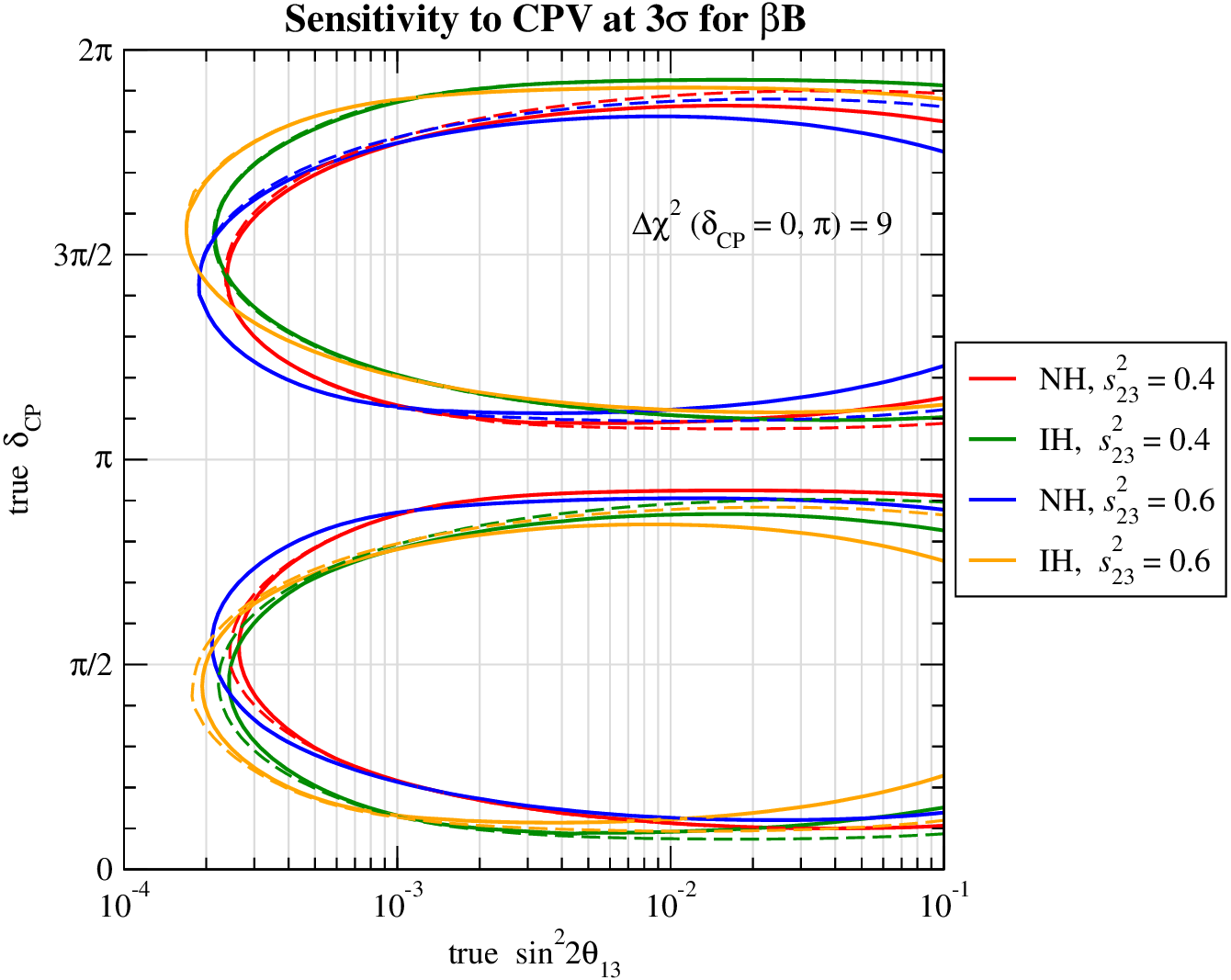}
   \mycaption{Impact of degeneracies on the CPV discovery potential
   for the \BB. We show the sensitivity to CPV at $3\sigma$
   $(\Delta\chi^2>9)$ computed for 4 different combinations of the
   true values of the hierarchy (NH or IH) and $\theta_{23}$
   ($\sin^2\theta_{23} = 0.4$ or $0.6$). Dashed curves are computed
   neglecting degeneracies in the fit. The running time is ($5\nu +
   5\bar\nu$)~yrs.
\label{fig:deltacp}}
\end{figure}

Finally, in Fig.~\ref{fig:deltacp} we illustrate the impact of
degeneracies, as well as the true hierarchy and \thetatt-octant on the
CPV sensitivity.  Curves of different colors correspond to the four
different choices for \sigdm\ and the \thetatt-octant of the true
parameters. For the solid curves the simulated data for each choice of
true \sigdm\ and \thetatt-octant are fitted by taking into account all
four degenerate solutions, i.e., also for the fit all four
combinations of \sigdm\ and \thetatt-octant are used. One observes
from the figure that the true hierarchy and octant have a rather small
impact on the \BB\ CPV sensitivity, in particular the sensitivity to
maximal CPV is completely independent.  The main effect of changing the
true hierarchy is to exchange the behavior between $0 < \delCP <
180^\circ$ and $180^\circ < \delCP < 360^\circ$. For $\stheta \lesssim
10^{-2}$ the sensitivity gets slightly worse if
$\thetatt^\mathrm{true} > \pi/4$ compared to $\thetatt^\mathrm{true} <
\pi/4$.

The dashed curves in Fig.~\ref{fig:deltacp} are computed without
taking into account the degeneracies, i.e., for each choice of true
\sigdm\ and \thetatt-octant the data are fitted only with this
particular choice. The effect of the degeneracies becomes visible for
large values of \thetaot. Note that this is just the region where they
can be reduced by a combined analysis with atmospheric neutrinos (see
Sec.~\ref{sec:atmospherics} or Ref.~\cite{Huber:2005ep}).

\section{Synergies provided by the CERN--MEMPHYS facilities}
\label{sec:synergies}

\subsection{Combining Beta Beam and Super Beam}
\label{sec:synergies-beams}

In this section we discuss synergies which emerge if both \BB\ and SPL
are available. The main difference between these two beams is the
different initial neutrino flavor,
$\stackrel{\scriptscriptstyle(-)}{\nu}_e$ for \BB\ and
$\stackrel{\scriptscriptstyle (-)}{\nu}_\mu$ for SPL. This implies
that at near detectors all relevant cross sections can be measured. In
particular, the near detector of the \BB\ will measure the cross
section for the SPL appearance search, and vice versa.
If both experiments run with neutrinos and antineutrinos all possible
transition probabilities are covered: $P_{\nu_e\to\nu_\mu}$,
$P_{\bar\nu_e\to\bar\nu_\mu}$, $P_{\nu_\mu\to\nu_e}$, and
$P_{\bar\nu_\mu\to\bar\nu_e}$. Together with the fact that matter
effects are very small because of the relatively short baseline, this
means that in addition to CP also direct tests of the T and CPT
symmetries are possible.

\begin{figure}[!t]
  \centering
   \includegraphics[width=0.9\textwidth]{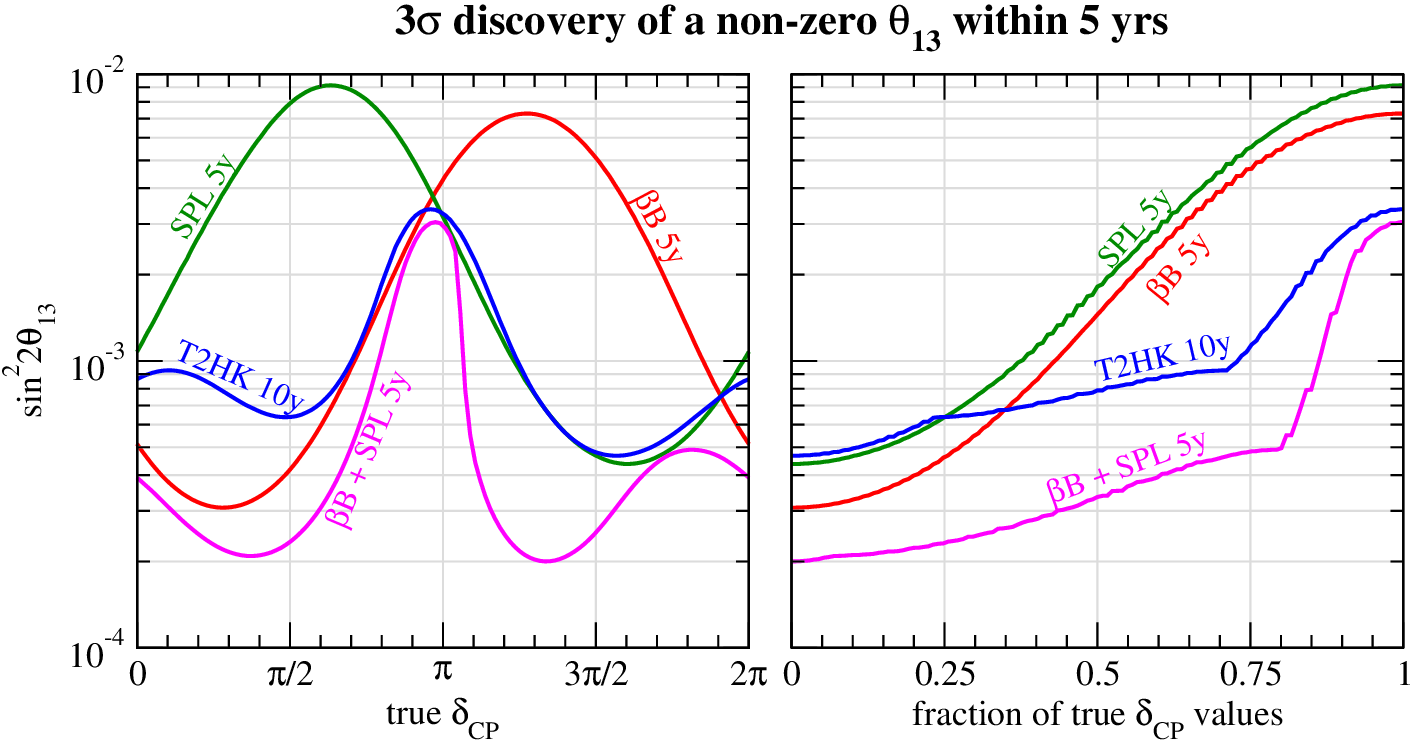}
   \mycaption{Discovery potential of a finite value of $\stheta$ at
   $3\sigma$ $(\Delta\chi^2>9)$ for 5~yrs neutrino data from
   \BB, SPL, and the combination of \BB\ + SPL compared to
   10~yrs data from T2HK (2~yrs neutrinos + 8~yrs antineutrinos).
   \label{fig:th13-5yrs}}
\end{figure}

\begin{figure}[!t]
  \centering
   \includegraphics[width=0.6\textwidth]{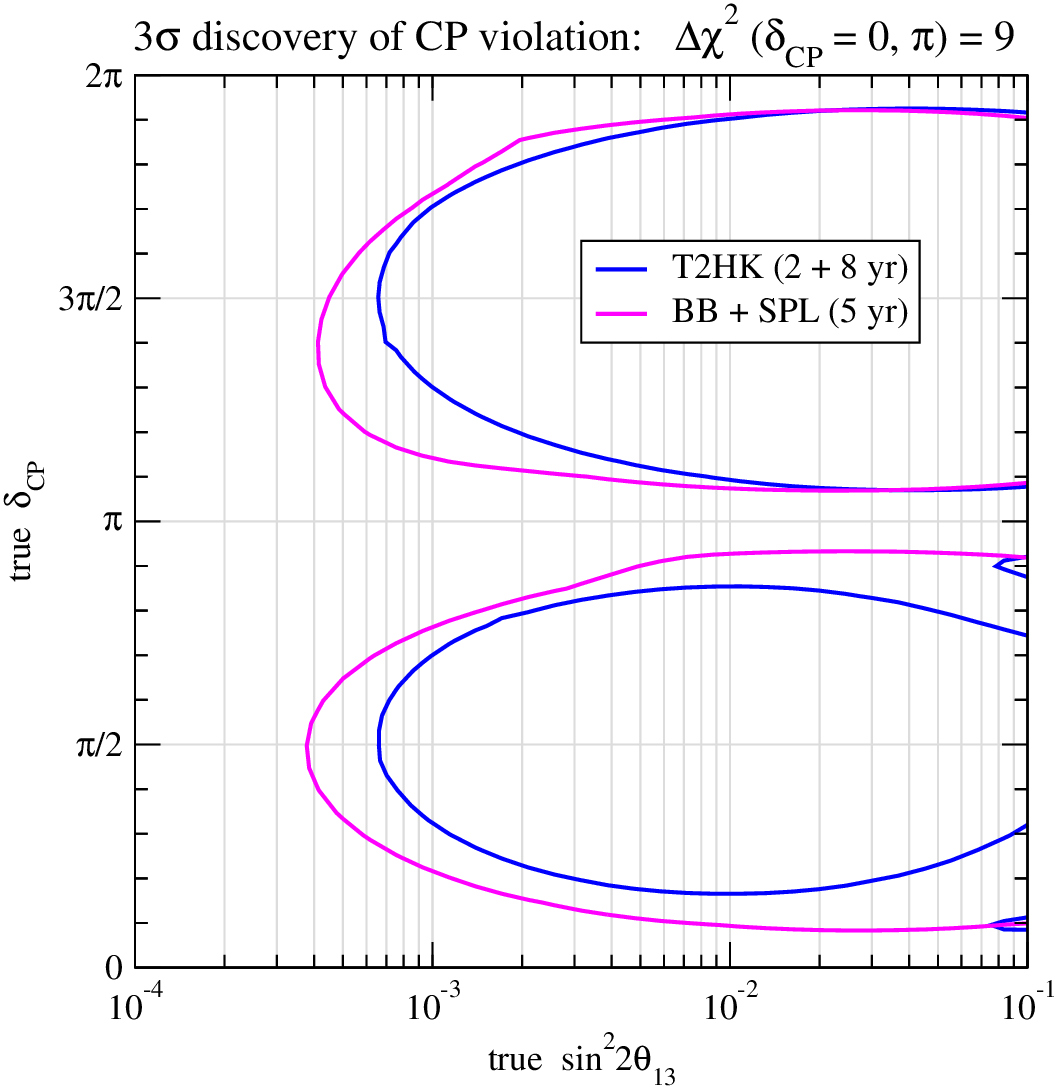}
   \mycaption{Sensitivity to CPV at $3\sigma$ $(\Delta\chi^2>9)$ for
   combining 5~yrs neutrino data from \BB\ and SPL compared to
   10~yrs data from T2HK (2~yrs neutrinos + 8~yrs antineutrinos).
   \label{fig:CP-5yrs}}
\end{figure}

However, if the CPT symmetry is assumed in principle all information
can be obtained just from neutrino data because of the relations
$P_{\bar\nu_e\to\bar\nu_\mu} = P_{\nu_\mu\to\nu_e}$ and
$P_{\bar\nu_\mu\to\bar\nu_e} = P_{\nu_e\to\nu_\mu}$. As mentioned
already in Sec.~\ref{sec:degeneracies} this implies that (time
consuming) antineutrino running can be avoided. We illustrate this
synergy in Figs.~\ref{fig:th13-5yrs} and \ref{fig:CP-5yrs}. In
Fig.~\ref{fig:th13-5yrs} we show the $\theta_{13}$ discovery potential
of 5 years of neutrino data from \BB\ and SPL. From the left panel the
complementarity of the two experiments is obvious, since each of them
is most sensitive in a different region of \delCP. (As expected from
general properties of the oscillation probabilities the sensitivity
curves of \BB\ and SPL are approximately related by the transformation
$\delCP \to 2\pi - \delCP$.) Combining these two data sets results in
a sensitivity slightly better than from 10 years (2$\nu$+8$\bar\nu$)
of T2HK data.
As visible in Fig.~\ref{fig:CP-5yrs} also for the CPV discovery this
synergy works and 5 years of neutrino data from \BB\ and SPL lead to a
similar sensitivity as 10 years of T2HK.

\subsection{Resolving degeneracies with atmospheric neutrinos}
\label{sec:atmospherics}

It was pointed out in Ref.~\cite{Huber:2005ep} that for LBL
experiments based on mega ton scale water \v{C}erenkov detectors data
from atmospheric neutrinos (ATM) provide an attractive method to
resolve degeneracies. Atmospheric neutrinos are sensitive to the
neutrino mass hierarchy if $\theta_{13}$ is sufficiently large due to
Earth matter effects, mainly in multi-GeV $e$-like
events~\cite{Petcov:1998su,Akhmedov:1998ui,Bernabeu:2003yp}. Moreover,
sub-GeV $e$-like events provide sensitivity to the octant of
$\theta_{23}$~\cite{Kim:1998bv,Peres:2003wd,Gonzalez-Garcia:2004cu}
due to oscillations with $\Delta m^2_{21}$ (see also
Ref.~\cite{Kajita} for a discussion of atmospheric neutrinos in the
context of Hyper-K). 
Following Ref.~\cite{Huber:2005ep} we investigate here the synergy
from a combination of LBL data from \BB\ and SPL with ATM data in the
MEMPHYS detector. Technical details are given in
Sec.~\ref{sec:atm-details}. 
 
The effect of degeneracies in LBL data has been discussed in
Sec.~\ref{sec:degeneracies}, see Figs.~\ref{fig:degeneracies} and
\ref{fig:degeneracies_SPL}. As discussed there, for given true
parameter values the data can be fitted with the wrong hierarchy
and/or with the wrong octant of $\theta_{23}$. Hence, from LBL data
alone the hierarchy and the octant cannot be determined and
ambiguities exist in the determination of $\theta_{13}$ and
$\delta_\mathrm{CP}$.
If the LBL data are combined with ATM data only the colored regions in
Fig.~\ref{fig:degeneracies} survive, i.e., in this particular example
for SPL and T2HK the degeneracies are completely lifted at 95\%~CL,
the mass hierarchy and the octant of $\theta_{23}$ can be identified,
and the ambiguities in $\theta_{13}$ and $\delta_\mathrm{CP}$ are
resolved. For the \BB\ an island corresponding to the wrong hierarchy
does survive at the 95\%~CL for 2~dof. Still, the solution with the
wrong sign of $\Delta m^2_{31}$ is disfavored with $\Delta\chi^2 =
5.1$ with respect to the true solution, which corresponds to
2.4$\sigma$ for 1~dof.
Let us note that in Fig.~\ref{fig:degeneracies} we have chosen a
favorable value of $\sin^2\theta_{23} = 0.6$; for values
$\sin^2\theta_{23} < 0.5$ in general the sensitivity of ATM data is
weaker~\cite{Huber:2005ep}.

\begin{figure}[!t]
\centering
  \includegraphics[width=0.9\textwidth]{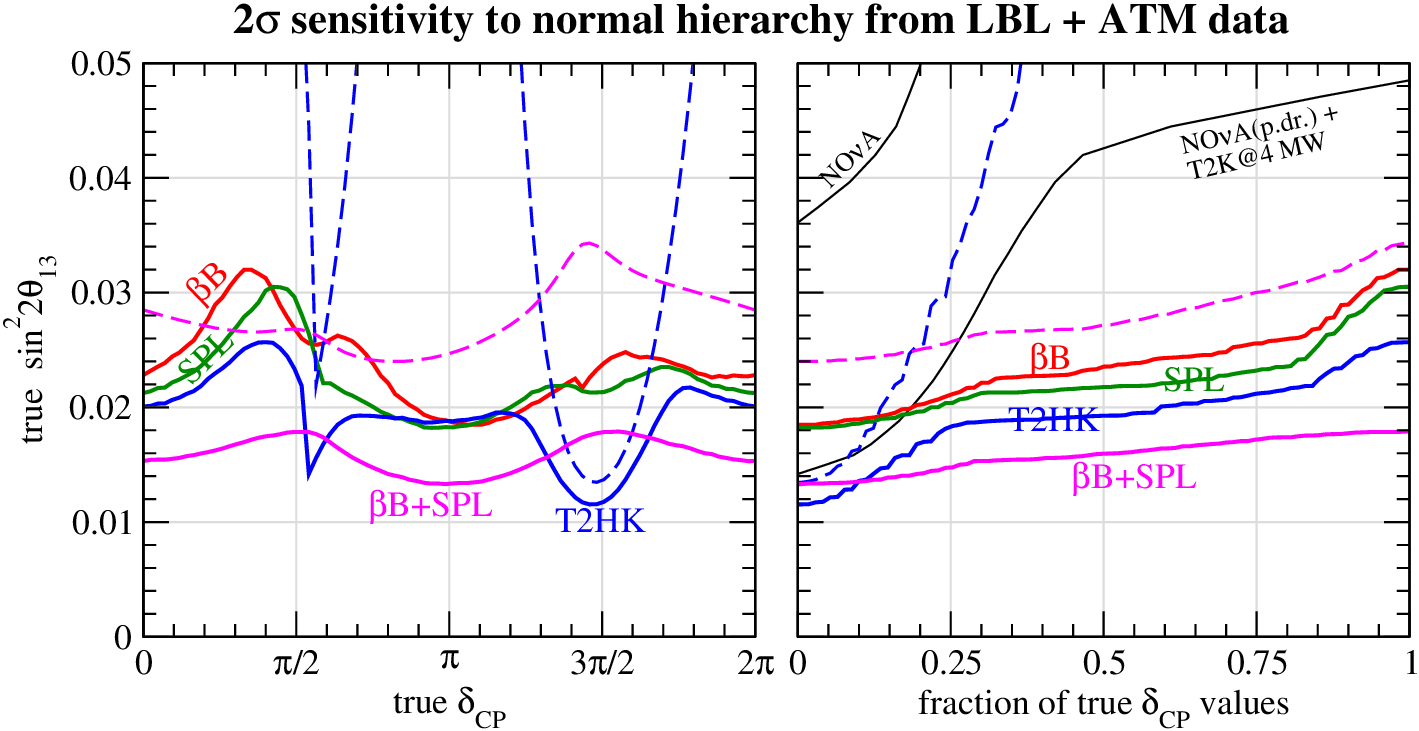}
  \mycaption{Sensitivity to the mass hierarchy at $2\sigma$
  $(\Delta\chi^2 = 4)$ as a function of the true values of
  $\sin^22\theta_{13}$ and $\delta_\mathrm{CP}$ (left), and the
  fraction of true values of $\delCP$ (right). The solid curves are
  the sensitivities from the combination of long-baseline and
  atmospheric neutrino data, the dashed curves correspond to
  long-baseline data only. The running time is ($5\nu + 5\bar\nu$)~yrs
  for \BB\ and ($2\nu + 8\bar\nu$)~yrs for the Super Beams. For
  comparison we show in the right panel also the sensitivities of
  NO$\nu$A and NO$\nu$A+T2K extracted from Fig.~13.14 of
  Ref.~\cite{Ayres:2004js}. For the curve labeled ``NO$\nu$A
  (p.dr.)+T2K@4~MW'' a proton driver has been assumed for NO$\nu$A and
  the T2K beam has been up-graded to 4~MW, see
  Ref.~\cite{Ayres:2004js} for details.}
  \label{fig:hierarchy}
\end{figure}

In Fig.~\ref{fig:hierarchy} we show how the combination of ATM+LBL
data leads to a non-trivial sensitivity to the neutrino mass
hierarchy, i.e.\ to the sign of $\Delta m^2_{31}$. For LBL data alone
(dashed curves) there is practically no sensitivity for the
CERN--MEMPHYS experiments (because of the very small matter effects
due to the relatively short baseline), and the sensitivity of T2HK
depends strongly on the true value of $\delta_\mathrm{CP}$. However,
by including data from atmospheric neutrinos (solid curves) the mass
hierarchy can be identified at $2\sigma$~CL provided
$\sin^22\theta_{13} \gtrsim 0.02-0.03$. As an example we have chosen
in that figure a true value of $\theta_{23} = \pi/4$. Generically the
hierarchy sensitivity increases with increasing $\theta_{23}$, see
Ref.~\cite{Huber:2005ep} for a detailed discussion.

The sensitivity to the neutrino mass hierarchy shown in
Fig.~\ref{fig:hierarchy} is significantly improved with respect to our
previous results obtained in Ref.~\cite{Huber:2005ep}. There are two
main reasons for this improved performance: First, we use now much
more bins in charged lepton energy for fully contained single-ring
events\footnote{The impact of energy binning on the hierarchy
determination with atmospheric neutrinos has been discussed recently
in Ref.~\cite{Petcov:2005rv} in the context of magnetized iron
detectors.}  (compare Sec.~\ref{sec:atm-details}), and second, we
implemented also information from multi-ring events. This latter point
is important since the relative contribution of neutrinos and
antineutrinos is different for single- and multi-ring
events. Therefore, combining single- and multi-ring data allows to
obtain a discrimination between neutrino and antineutrino events on a
statistical basis. This in turn contains crucial information on the
hierarchy, since the matter enhancement is visible either in neutrinos
or antineutrinos, depending on the hierarchy.

Although \BB\ and SPL alone have no sensitivity to the hierarchy at
all, we find that the combination of them does provide rather good
sensitivity even without atmospheric data. The reason for this
interesting effect is the following. Because of the rather short
baseline the matter effect is too small to distinguish between NH and
IH given only neutrino and antineutrino information in one channel.
However, the tiny matter effect suffices to move the hierarchy
degenerate solution to slightly different locations in the ($\stheta$,
$\delCP$) plane for the $\stackrel{\scriptscriptstyle(-)}{\nu}_e \to
\stackrel{\scriptscriptstyle (-)}{\nu}_\mu$ (\BB) and
$\stackrel{\scriptscriptstyle(-)}{\nu}_\mu \to
\stackrel{\scriptscriptstyle (-)}{\nu}_e$ (SPL) channels (compare
Fig.~\ref{fig:degeneracies}). Hence, if all four CP and T conjugate
channels are available (as it is the case for the \BB+SPL combination)
already the small matter effect picked up along the 130~km
CERN--MEMPHYS distance provides sensitivity to the mass hierarchy for
$\sin^22\theta_{13} \gtrsim 0.03$, or $\sin^22\theta_{13} \gtrsim 0.015$
if also atmospheric neutrino data is included.

For comparison we show in the right panel of Fig.~\ref{fig:hierarchy}
also the sensitivity of the NO$\nu$A~\cite{Ayres:2004js} experiment,
and of NO$\nu$A+T2K, where in the second case a beam upgrade by a
proton driver has been assumed for NO$\nu$A, and for T2K the
Super-Kamiokande detector has been used but the beam intensity has
been increased by assuming 4~MW power. More details on these
sensitivities can be found in Ref.~\cite{Ayres:2004js}.
Let us note that in general LBL experiments with two detectors and/or
very long baselines ($\gtrsim 1000$~km) are a competitive method to
atmospheric neutrinos for the hierarchy determination, see, e.g.,
Refs.~\cite{Ishitsuka:2005qi,Kajita:2006bt,MenaRequejo:2005hn,Hagiwara:2005pe,Barger:2006vy}
for recent analyses. In particular, in case of the T2KK extension of
the T2HK experiment, the very long baseline to Korea allows for a
determination of the mass hierarchy down to $\sin^22\theta_{13}
\gtrsim 0.02$ ($2\sigma$) without using atmospheric
neutrinos~\cite{Kajita:2006bt}.
We mention also the possibility to determine the neutrino mass
hierarchy by using neutrino events from a galactic Super Nova
explosion in mega ton \v{C}erenkov detectors such as MEMPHYS, see,
e.g., Ref.~\cite{Kachelriess:2004vs}.

\begin{figure}[!t]
\centering
  \includegraphics[width=0.55\textwidth]{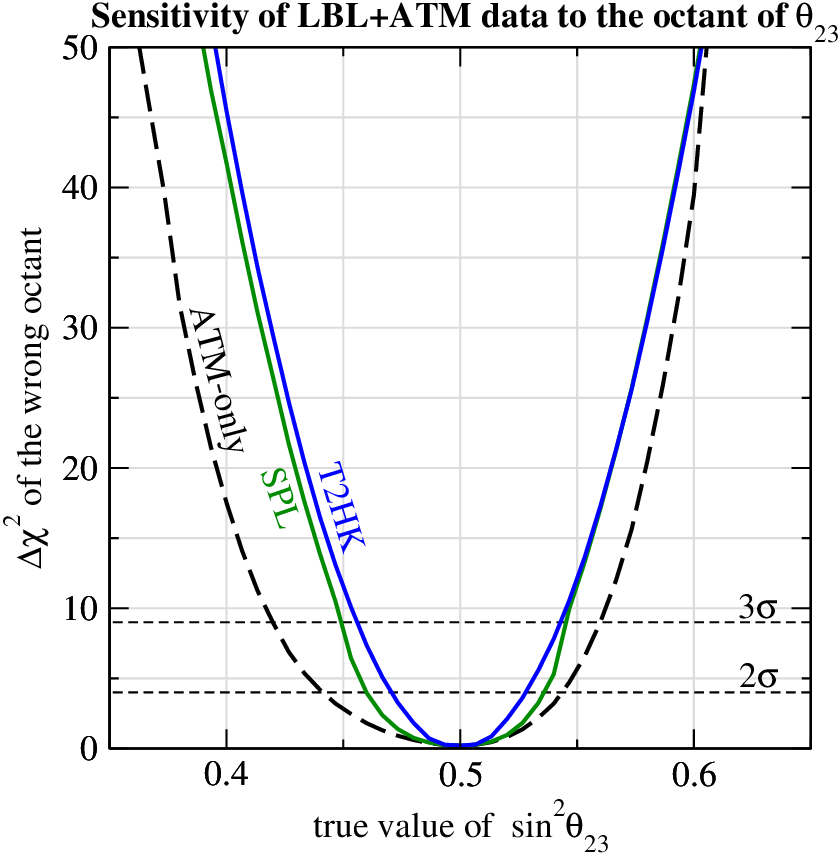}
  \mycaption{$\Delta\chi^2$ of the solution with the wrong octant of
  $\theta_{23}$ as a function of the true value of
  $\sin^2\theta_{23}$. We have assumed a true value of $\theta_{13} =
  0$, and the running time is ($2\nu + 8\bar\nu$)~yrs.}
  \label{fig:octant}
\end{figure}

Fig.~\ref{fig:octant} shows the potential of ATM+LBL data to exclude
the octant degenerate solution. Since this effect is based mainly on
oscillations with $\Delta m^2_{21}$ there is very good sensitivity
even for $\theta_{13} = 0$; a finite value of $\theta_{13}$ in general
improves the sensitivity~\cite{Huber:2005ep}.  From the figure one can
read off that atmospheric data alone can can resolve the correct
octant at $3\sigma$ if $|\sin^2\theta_{23} - 0.5| \gtrsim 0.085$. If
atmospheric data is combined with the LBL data from SPL or T2HK there
is sensitivity to the octant for $|\sin^2\theta_{23} - 0.5| \gtrsim
0.05$. The improvement of the octant sensitivity with respect to
previous analyses~\cite{Huber:2005ep,Gonzalez-Garcia:2004cu} follows
from changes in the analysis of sub-GeV atmospheric events, where now
three bins in lepton momentum are used instead of one. Note that since
in Fig.~\ref{fig:octant} we have assumed a true value of $\theta_{13}
= 0$, combining the \BB\ with ATM does not improve the sensitivity
with respect to atmospheric data alone. 
We note that the T2KK configuration provides also some sensitivity to
the octant of $\theta_{23}$ without referring to atmospheric
neutrinos. In Ref.~\cite{Kajita:2006bt} it was found that for
$|\sin^2\theta_{23} - 0.5| \gtrsim 0.12$ the octant can be identified
in T2KK at $3\sigma$.

\section{Summary}
\label{sec:conclusions}

In this work we have studied the physics potential of the
CERN--MEMPHYS neutrino oscillation project. We consider a Beta Beam
(\BB) with $\gamma = 100$ for the stored ions, where existing
facilities at CERN can be used optimally, and a Super Beam based on an
optimized Super Proton Linac (SPL) with a beam energy of 3.5~GeV and
4~MW power. As target we assume the MEMPHYS detector, a 440~kt water
\v{C}erenkov detector at Fr\'ejus, at a distance of 130~km from
CERN. The main characteristics of the experiments are summarized in
Tab.~\ref{tab:setups}.
The adopted neutrino fluxes are based on realistic calculations of ion
production and storage for the \BB, and a full simulation of the beam
line (particle production and decay of secondaries) for SPL. Special
care has be given to the issue of backgrounds, which we include by
means of detailed event simulations and applying Super-Kamiokande particle
identification algorithms.

The physics potential of the \BB\ and SPL experiments in terms of
$\theta_{13}$ discovery reach and sensitivity to CP violation has been
addressed where parameter degeneracies are fully taken into account.
The main results on these performance indicators are summarized in
Figs.~\ref{fig:th13} and \ref{fig:CPV}.
We obtain a guaranteed discovery reach of $\stheta \simeq 5\times
10^{-3}$ at $3\sigma$, irrespective of the actual value of \delCP. For
certain values of \delCP\ the sensitivity is significantly improved,
and for \BB\ (SPL) discovery limits around $\stheta \simeq 3\,(10)
\times 10^{-4}$ are possible for a large fraction of all possible
values of \delCP.
Maximal CP violation (for $\delCP^\mathrm{true} = \pi/2, \, 3\pi/2$)
can be discovered at $3\sigma$ down to $\stheta \simeq 2\, (9)\times
10^{-4}$ for \BB\ (SPL), whereas the best sensitivity to CP violation
is obtained for $\stheta \sim 10^{-2}$: For $\stheta = 10^{-2}$ CP
violation can be established at $3\sigma$ for 78\% (73\%) of all
possible true values of \delCP\ for \BB\ (SPL).
We stress that the \BB\ performance in general depends crucially on
the number of ion decays per year.
The impact of the value of systematical uncertainties on signal and
background on our results is discussed.
The \BB\ and SPL sensitivities are compared to the ones of the
phase~II of the T2K experiment in Japan (T2HK), which is a competing
proposal of similar size and timescale. In general we obtain rather
similar sensitivities for T2HK and SPL, and hence the CERN--MEMPHYS
experiments provide a viable alternative to T2HK. We find that \BB\
and SPL are less sensitive to systematical errors, whereas the
sensitivity of T2HK crucially depends on the systematical error on the
background.\footnote{Let us note that in the present study we have not
considered the recent ``T2KK'' proposal~\cite{Ishitsuka:2005qi}, where
one half of the Hyper-K detector mass is at Kamioka and the second
half in Korea. For such a setup our results do not apply and
especially the conclusion on systematical errors may be different.}

Assuming that both \BB\ and SPL experiments are available, we point
out that one can benefit from the different oscillation channels
$\nu_e\to\nu_\mu$ for \BB\ and $\nu_\mu\to\nu_e$ for SPL, since by the
combination of these channels the time intensive antineutrino
measurements can be avoided. We show that 5 years of neutrino data from
\BB\ and SPL lead to similar results as 2 years of neutrino plus 8
years of antineutrino data from T2HK.
Furthermore, we discuss the use of atmospheric neutrinos in the
MEMPHYS detector to resolve parameter degeneracies in the
long-baseline data. This effect leads to a sensitivity to the neutrino
mass hierarchy at $2\sigma$~CL for $\sin^22\theta_{13} \gtrsim
0.025$ for \BB\ and SPL, although these experiments alone (without
atmospheric data) have no sensitivity at all. The optimal hierarchy
sensitivity is obtained from combining \BB+SPL+atmospheric data.
Furthermore, the combination of atmospheric data with a Super Beam
provides a possibility to determine the octant of $\theta_{23}$.

To conclude, we have shown that the CERN--MEMPHYS neutrino oscillation
project based on a Beta Beam and/or a Super Beam plus a mega ton scale
water \v{C}erenkov detector offers interesting and competitive physics
possibilities and is worth to be considered as a serious option in
the worldwide process of identifying future high precision neutrino
oscillation facilities~\cite{ISSpage}.

\subsection*{Acknowledgment}

We thank J.~Argyriades for communication on the Super-K atmospheric
neutrino analysis, A.~Cazes for his work on the SPL simulation, and
P.~Huber for his patience in answering questions concerning the use of
GLoBES. Furthermore, we thank D.~Casper for the help in installing and
running Nuance, and E.~Couce for discussions about the \BB\
backgrounds.  T.S.\ is supported by the $6^\mathrm{th}$~Framework
Program of the European Community under a Marie Curie Intra-European
Fellowship.


\end{document}